# Short-range tests of the equivalence principle

G. L. Smith, C. D. Hoyle, J. H. Gundlach, E. G. Adelberger, B. R. Heckel, and H. E. Swanson
*Department of Physics, University of Washington, Seattle, Washington 98195*

We tested the equivalence principle at short length scales by rotating a 3 ton $^{238}$U attractor around a compact torsion balance containing Cu and Pb test bodies. The observed differential acceleration of the test bodies toward the attractor, $a_{\mathrm{Cu}}-a_{\mathrm{Pb}}=(1.0\pm 2.8)\times 10^{-13}$ cm/s$^2$, should be compared to the corresponding gravitational acceleration of $9.2\times 10^{-5}$ cm/s$^2$. Our results set new constraints on equivalence-principle violating interactions with Yukawa ranges down to 1 cm, and improve by substantial factors existing limits for ranges between 10 km and 1000 km. Our data also set strong constraints on certain power-law potentials that can arise from two-boson exchange processes.

PACS number(s): 04.80.−y, 04.90.+e, 14.80.−j

## I. MOTIVATION

The most precise tests of the equivalence principle, the underlying symmetry of general relativity, are made by comparing the accelerations of test bodies toward a massive attractor. The equivalence principle, which states that gravitation is locally equivalent to an acceleration of the reference frame, predicts that these gravitational accelerations will be identical if tidal forces can be neglected. The classic Princeton [1] and Moscow [2] equivalence-principle tests compared the accelerations of Al and Au or Pt test bodies toward the Sun, implicitly assuming that any violation of the equivalence principle, or more precisely the universality of free fall (UFF), would have an infinite range. This was a natural assumption in a classical context where the equivalence principle can be viewed in Newtonian terms; i.e., are gravitational and inertial masses identical? However, in a quantum context it is natural to view equivalence principle tests as probes for new Yukawa interactions arising from exchange of exotic, low-mass bosons. In this context the tests should be sensitive to UFF violation over a broad variety of Yukawa length scales and should utilize a variety of test bodies whose relevant properties, such as binding energy per unit mass, atomic charge $Z$, neutron-to-proton ratio $N/Z$, etc., differ by the greatest practical amount. Such UFF tests are powerful probes of new physics because interactions arising from exchange of scalar or vector bosons generically violate the UFF (see Ref. [3]).

We have undertaken a series of experiments with the goal of testing the UFF with as much generality as is practical. In designing these tests, we imagine UFF-violating interactions arising from the exchange of scalar or vector bosons of mass $m_{\mathrm{b}}$. These will produce a potential between point bodies with a Yukawa form

$$V_{12}(r)=\mp\frac{f^2}{4\pi}\tilde{q}_1\tilde{q}_2\frac{e^{-r/\lambda}}{r}, \tag{1}$$

where $f$ is a coupling constant, $\tilde{q}$ is the Yukawa charge of a body (not to be confused with its electrical charge), $\lambda=\hbar/m_{\mathrm{b}}c$ is the interaction range, and the − and + signs correspond to interactions mediated by scalar or vector bosons respectively. We make no *a priori* assumptions concerning $\tilde{q},\lambda$, or the scalar or vector nature of the interaction. The interaction in Eq. (1) would lead to a differential acceleration of test bodies 1 and 2 toward an attractor A:

$$\frac{\Delta\vec{a}}{g'}\equiv\frac{\vec{a}_1-\vec{a}_2}{g'}=\alpha\left[\left(\frac{\tilde{q}}{\mu}\right)_1-\left(\frac{\tilde{q}}{\mu}\right)_2\right]\left(\frac{\tilde{q}}{\mu}\right)_{\mathrm{A}}\vec{I}_{\mathrm{A}}(\lambda), \tag{2}$$

where $g'$ is the gravitational acceleration toward A, $\alpha=\pm f^2/(4\pi G u^2)$ is dimensionless, $u$ is the atomic mass unit, $\mu$ refers to the mass in units of $u$, and $\vec{I}_{\mathrm{A}}$ is a dimensionless integral over the test-body and attractor geometries.

We recently reported sensitive torsion-balance tests of the UFF for Be, Al, Si/Al and Cu test bodies attracted toward the Earth, the Sun, and the dark matter in our galaxy [4]. This paper describes our exploration of the short-range regime, where we compared the accelerations of Cu and Pb toward a rotating 3 ton uranium attractor. The experimental results are based on two data sets, taken in 1996 and 1997–1998 respectively. A brief account of the 1996 work, which constituted the thesis project of Smith [5] has appeared in Letter form [6]. This paper presents a complete description of the apparatus and experimental techniques and an improved analysis of the 1996 data, as well as the 1997–1998 results.

We undertook this work for the following reasons.

(1) Previous constraints on UFF-violating interactions with $\lambda<1$ m were relatively weak. However, the regime 1 cm$\leqslant\lambda\leqslant$1 m is an interesting one to explore. On general grounds, there may be some significance to the range $\lambda\approx 30$ cm which corresponds to an exchanged boson with a mass $m_{\mathrm{b}}=M_{\mathrm{Z}}^2/M_{\mathrm{P}}$, where $M_{\mathrm{P}}$ is the Planck mass and $M_{\mathrm{Z}}\approx 90$ GeV is a mass scale associated with the standard model of elementary particles. More particularly, superstring theories with low-energy supersymmetry breaking predict scalar particles (moduli) that could lead to UFF-violating forces with $\lambda\lesssim 1$ cm [7].

(2) Although our torsion-balance tests [4] using Earth and its local topography as the attractor had good sensitivity to UFF-violating interactions with 1 m$\leqslant\lambda\leqslant\infty$, we did not quote constraints for ranges between $\lambda=10$ km and $\lambda=1000$ km. This gap occurred because torsion balances are sensitive only to the *horizontal* components of acceleration. For ranges between 1 m and 10 km, the horizontal compo-nent of the Earth's Yukawa field is adequately determined by local topography [4]. For $\lambda \geqslant 1000$ km, the horizontal field can be computed by modeling the Earth as a fluid in equilibrium under gravitational and centrifugal forces [4]. However, neither local topography nor global models adequately describe the horizontal Yukawa field in the "gap" region where 10 km$<\lambda<$1000 km. At these length scales the horizontal field is dominated by poorly known subsurface topography (the mountains, for example, have low density and float on higher density substrata).

(3) Equivalence-principle tests that use the Earth as an attractor necessarily have low sensitity to UFF-violating interactions with $\tilde{q} \propto N-Z$ because the Earth contains nearly equal numbers of neutrons and protons. In fact, there was an unconfirmed suggestion that such an interaction may exist [8].

Our new test used a $^{238}$U attractor because its density $\rho \approx 18.4$ g/cm$^3$ [9] is very high, its neutron excess $(N-Z)/(N+Z)=0.23$ differs substantially from the much smaller terrestrial value, and we can readily compute $\vec{I}_{\rm A}(\lambda)$ for any $\lambda$. The high density allowed us to design an instrument with very close geometry and good sensitivity to Yukawa interactions with ranges down to 1 cm, corresponding to exchanged bosons with masses up to $2\times10^{-5}$ eV. In addition, the attractor was sufficiently massive that we could improve on the best previous constraints in the "gap region" that came from Galileo-type comparisons of vertical accelerations [10,11].

## II. ROT-WASH TORSION BALANCE

### A. Overview

The Rot-Wash instrument, shown schematically in Fig. 1, consisted of a massive, specially shaped uranium attractor that revolved slowly and uniformly around a compact, stationary torsion balance. The balance contained a highly symmetric torsion pendulum, shown in Fig. 2, holding 2 Cu and 2 Pb test bodies configured as a composition dipole. The pendulum was freely suspended by a fine tungsten fiber inside a vessel whose pressure was maintained at $1\times10^{-6}$ Torr by a small ion pump. External torques on the pendulum were measured by monitoring its angular deflection (twist) with an autocollimator. The torsion balance instrument was mounted on a triangular granite table supported by three air-table legs [12]. The air pressure in the legs could be controlled by non-contacting position sensors mounted above each leg. We did not find it necessary to float the granite table during normal operation, but the ability to float the table in a well-controlled way proved very useful when measuring the "sidepull effect" discussed in Sec. IV B 2.

### B. UFF-violating signal

A UFF-violating interaction would apply a torque on the pendulum that varied as the harmonic function of the azimuthal angle, $\phi$, between the composition dipole and the uranium attractor:

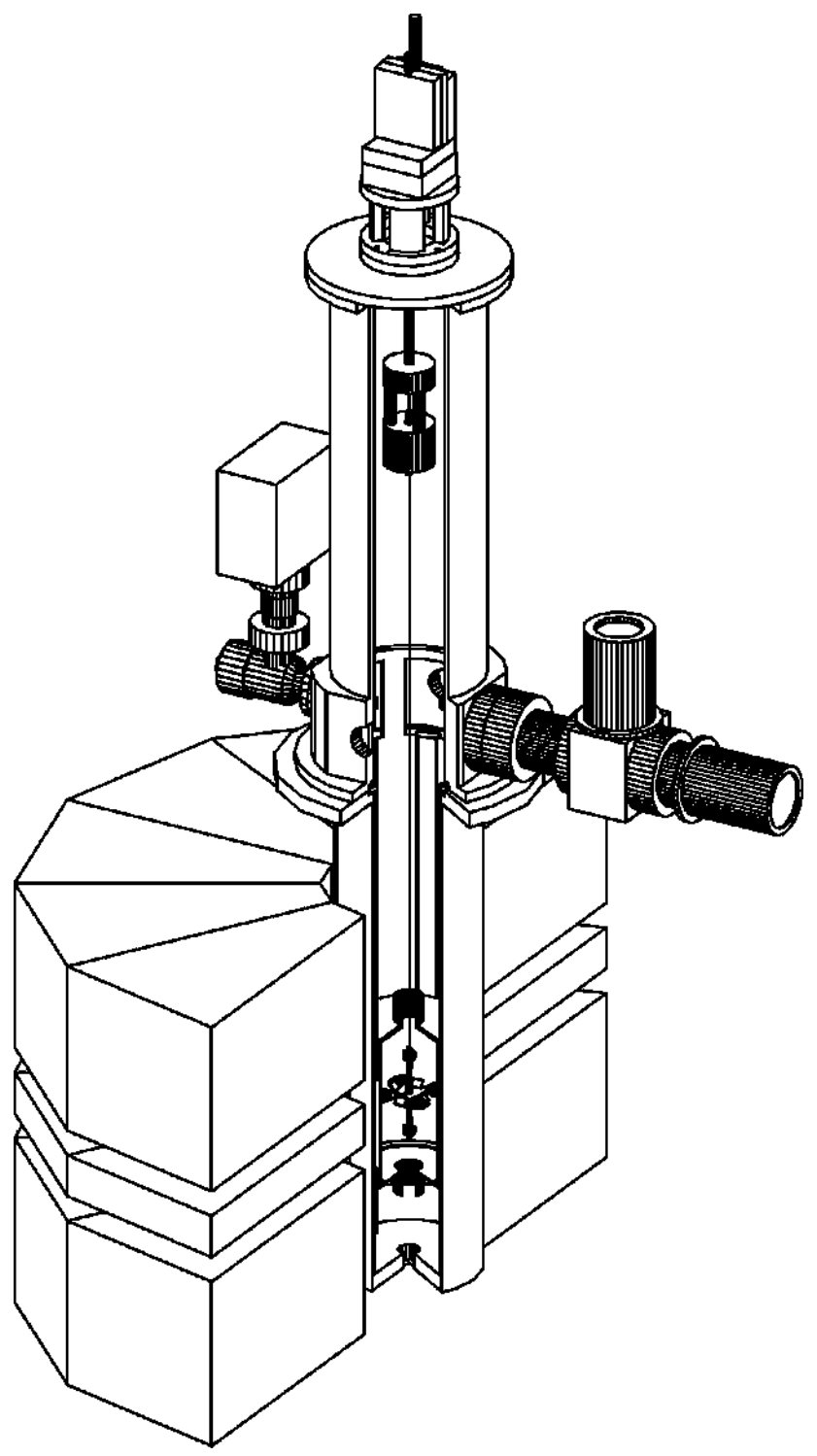

FIG. 1. Scale drawing of the Rot-Wash instrument; the vacuum vessel has been cut away to show the damper, pendulum and inner magnetic shield. The rotating 2630 kg uranium attractor was counterbalanced by 820 kg of Pb, shown in Fig. 3, to prevent the laboratory floor from tilting as the attractor revolved. The attractor turntable, the counterbalance, and the small Pb blocks that minimized the attractor's stray $Q_{21}$ and $Q_{31}$ fields are omitted for clarity. The instrument was surrounded by a constant-temperature enclosure and the entire apparatus resided inside a temperature-controlled shed.

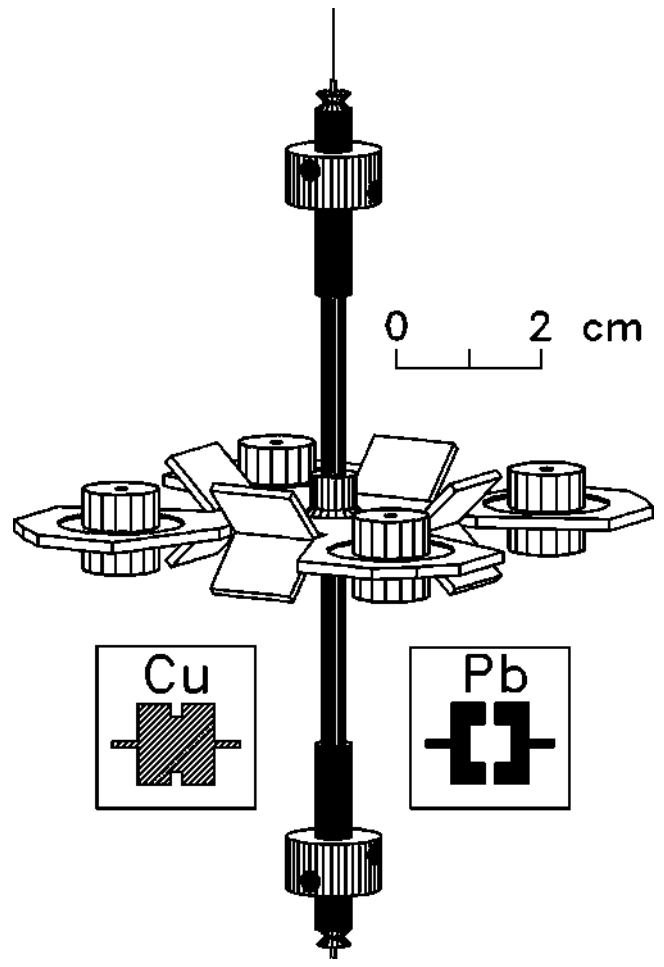

FIG. 2. The torsion pendulum. The masses on the ends of the vertical shaft made the $q_{20}$ moment of the entire pendulum vanish. Small screws in these masses were adjusted to reduce the stray $q_{21}$ and $q_{31}$ moments of the pendulum. The inset shows cross sections of the cylindrically symmetric test bodies.

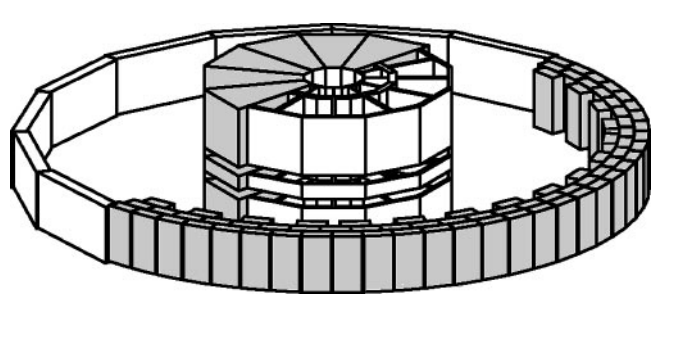

FIG. 3. The rotating attractor in its normal *state*. The uranium and the lead counterbalance are shaded. The pseudo-attractor and pseudo-counterbalance, made from hollow aluminum structures and blocks of foam, respectively, are shown unshaded.

$$T_{\mathrm{Y}}=8\pi G\alpha\sum_{l}\frac{1}{2l+1}\sum_{m=1}^{l}m\mathrm{Re}[e^{-im\phi}q_{lm}^{5}(\lambda)Q_{lm}^{5}(\lambda)],\tag{3}$$

where $q_{lm}^5$ and $Q_{lm}^5$ [defined in Eqs. (15) and (16) of Ref. [13]] are inner and outer Yukawa spherical multipole moments of the Yukawa charge. The dominant $m=1(1\omega)$ component of this torque can be written as

$$T_{\mathrm{Y}}^{(1)}=\alpha\Delta\left(\frac{\tilde{q}}{\mu}\right)\langle(\tilde{q}/\mu)_{\mathrm{A}}I_{\mathrm{A}}(\lambda)\rangle Mdg'\sin\phi,\tag{4}$$

where $\Delta(\tilde{q}/\mu)=(\tilde{q}/\mu)_{\mathrm{Cu}}-(\tilde{q}/\mu)_{\mathrm{Pb}}$, $\langle(\tilde{q}/\mu)_{\mathrm{A}}I_{\mathrm{A}}(\lambda)\rangle$ is an appropriate average over attractor materials, $M=9.980$ g is the mass of a single test body, $d=4.31$ cm is the distance between centers of adjacent test bodies, $g'=9.2\times10^{-5}$ cm/s$^2$ is the gravitational acceleration toward the attractor, and

$$I_{\mathrm{A}}(\lambda)=\sum_{l=1}^{\infty}\frac{3}{2l+1}\frac{\mathrm{Re}[q_{l1}^{5}(\lambda)Q_{l1}^{5}(\lambda)]}{\mathrm{Re}[q_{11}^{5}(\infty)Q_{11}^{5}(\infty)]}.\tag{5}$$

For our instrument terms in Eq. (5) with $l>3$ are completely negligible.

Because of the close geometry of our instrument we had to devote much effort to minimizing spurious torques from gravity gradients. These torques are proportional to products of the $q_{lm}$ inner gravitational multipole moments of the pendulum and the $Q_{lm}$ outer gravitational multipole moments of the attractor defined in Eqs. (10) and (11) of Ref. [4]:

$$T_{\mathrm{g}}=8\pi G\sum_{l=2}^{\infty}\frac{1}{2l+1}\sum_{m=1}^{l}m[\mathrm{Re}(q_{lm}Q_{lm})\sin m\phi+\mathrm{Im}(q_{lm}Q_{lm})\cos m\phi].\tag{6}$$

The sum in Eq. (6) begins with $l=2$ because a pendulum suspended from a perfectly flexible fiber has vanishing $l=1$ moments. We paid special attention to minimizing the $m=1$ gravitational torques because they have the same $\phi$ dependence as a true UFF-violating effect.

### C. Attractor

The high-purity, depleted uranium attractor, shown in Fig. 3, had a total mass of 2630 kg and consisted of 108 2.6-cm-thick trapezoidal blocks stacked in 6 columns. These formed an essentially annular shape with effective inner and outer radii of 10.2 cm and 44.6 cm, respectively. The attractor spanned 180° of azimuth and was divided into 3 segments by two horizontal gaps. The gaps minimized the $Q_{31}$ and $Q_{51}$ fields. The $Q_{21},Q_{41},Q_{61},\ldots$ fields nominally vanished by reflection symmetry about the horizontal midplane. The 3 attractor segments were separated by non-magnetic bearings that allowed us to rotate the upper 2 segments with respect to the lowest segment as shown in Fig. 4. These alternate attractor *states* produced the large $Q_{21}$ or $Q_{31}$ fields used to measure the stray $q_{lm}$ moments of the pendulum as discussed in Sec. III C 3.

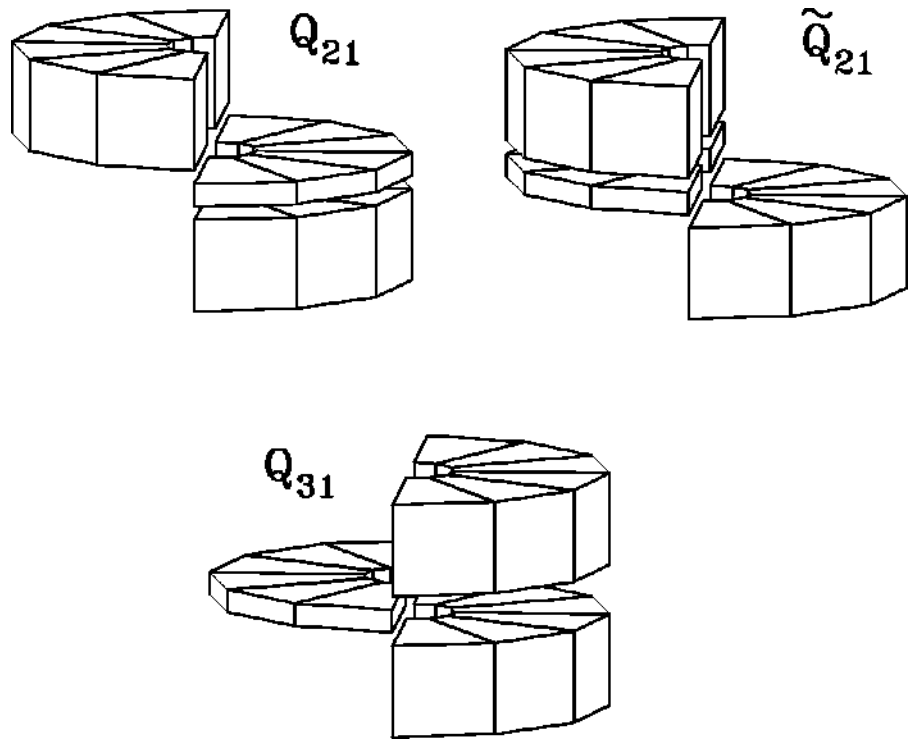

FIG. 4. Attractor *states* used to measure stray pendulum moments. The $q_{21}$ moment was measured by averaging data taken with the $Q_{21}$ and $\tilde{Q}_{21}$ *states*. The $q_{31}$ moment was found using the $Q_{31}$ *state*.

Considerable effort was spent in tuning the attractor to reduce stray gravity gradients. Because the commercially fabricated uranium blocks failed to meet our specified tolerances, we were forced to weigh and size each block and assemble them in a pattern that minimized the stray gravity gradients. However, the residual gradients (measured using gradiometers discussed in Sec. III B 2) were still unacceptably large. These stray gradients were reduced by adding small lead blocks in appropriate places.

The uranium attractor was mounted on a turntable constructed from non-magnetic materials except for a central ball bearing that was de-Gaussed before installation. The turntable angle was digitized by an encoder that gave 270 000 pulses/revolution. The weight of the attractor was counterbalanced by 820 kg of lead placed on the outer rim of the turntable; this prevented the floor from tilting as the attractor rotated. The counterbalance was tuned using an electronic level on one of the support legs of the turntable; an imbalance of 1 kg at the outer edge of the turntable could easily be detected. To prevent air currents in the apparatus from being affected by the attractor rotation, we installed a ''pseudo-attractor'' and a ''pseudo-counterbalance,'' constructed from impermeable, low-density materials, which effectively made the attractor azimuthally symmetric to the air currents.

### D. Torsion pendulum

The pendulum, shown in Fig. 2, held 2 passive compensator masses and 4 test bodies in a Be ''tray.'' It had a total mass of 59.26 g. The leading $m=1$ moment of the entire pendulum nominally occurred in $l=7$ multipole order. The odd-$l$ moments of the pendulum nominally vanished by reflection symmetry about the horizontal midplane. The compensator masses minimized the $q_{20}$ moment of the pendulum, preventing it from acquiring a significant $q_{21}$ moment from ''tip'' caused by small misalignments of the fiber attachment. However, construction imperfections produced stray moments in $l=2$ and higher order. The pendulum geometry was checked with a measuring microscope and 3 significant flaws were observed:

(1) The central ''axle'' was bent in the middle by 20 $\mu$m. This produced $q_{31}$ and, less importantly, $q_{51}$ moments.

(2) The normal to the plane of the 4 ''arms'' of the pendulum tray differed by 1 mrad from the axis of the ''axle.'' This produced $q_{21}$, and less importantly, $q_{41}$ moments.

(3) The fiber was not exactly centered on the pendulum axis. This tipped the pendulum, inducing [see Eq. (14) of Ref. [4]] $q_{41}$ and $q_{43}$ moments from the non-zero $q_{40}$ and $q_{44}$ moments, respectively, of the as-designed pendulum.

The pendulum's stray $l=2$ and $l=3$ moments were minimized by configuring the attractor to produce large $Q_{21}$ and $Q_{31}$ fields and then adjusting small trim screws in the passive compensator masses to tune out the $m=1$ torque on the pendulum.

The odd-$l$ intrinsic moments of the test bodies themselves vanished by symmetry, while the $l=2$ and $l=4$ moments vanished by design. All test bodies had masses of 9.980 g and nominally identical external dimensions. The solid Cu bodies were machined from 99.996% pure material. The Pb bodies were made from a 92% Pb/7.75% Sb/0.25% Sn alloy for good machinability, and contained machined cavities to account for their greater density.

The test bodies were held with their centers 3.05 cm from the pendulum axis by machined recesses in the pendulum tray. This allowed us to interchange the test bodies and reverse the composition dipole without altering the rest of the instrument. The composition dipole along with the rest of the pendulum and suspension fiber could also be reversed with respect to the remainder of the apparatus by rotating the fiber's upper attachment by 180°.

Four 90° mirrors were attached to the pendulum tray, any one of which could be used to monitor the pendulum twist. The entire pendulum including the mirrors was coated with Au and suspended inside a Au-coated magnetic shield by a $\sim$80 cm long, 20 $\mu$m diameter Au-coated W fiber [14]. The upper end of the fiber was attached to a cylindrically symmetric eddy-current damper that damped the ''swing,'' ''guitar string,'' and ''wobble'' modes of the pendulum in times short compared to the $\tau_0=747.4$ s torsional period, while having little effect on the torsional mode whose amplitude damping time, $\tau_{\rm d}=5\times10^5$ s, corresponded to a quality factor $Q_0=\pi\tau_{\rm d}/\tau_0=2100$ (see Fig. 5). The upper part of the damper was attached to a 2 cm long, 75 $\mu$m diameter W prehanger wire. This prehanger ensured that the upper attachment of the suspension fiber was always essentially vertical. The prehanger wire was supported by a translation-rotation stage that oriented the pendulum so that the autocollimator beam could strike the center of any of its 4 mirrors. The translation-rotation stage was electrically insulated from the rest of the apparatus. In normal operation, the pendulum was grounded to the apparatus by an external connection. However, the insulation proved useful in ''cooling'' the pendulum before taking data; by applying properly timed 100 V pulses to the pendulum we could produce electrostatic torques that reduced the pendulum's initial torsional amplitude close to the thermal value, $\theta^{\rm th}_{\rm rms}=\sqrt{kT/\kappa}=1.1$ $\mu$rad, where $\kappa=.031$ erg/rad (see Sec. III A 2) is the torsional constant of the fiber.

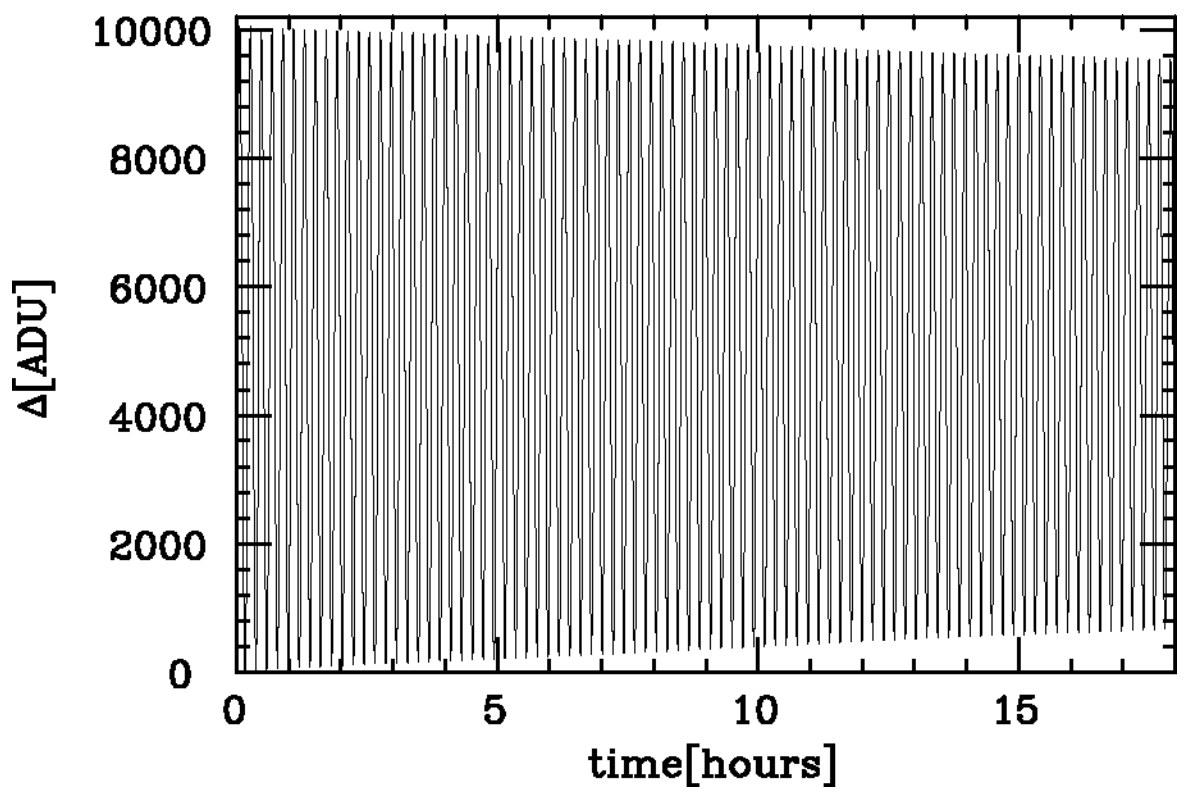

FIG. 5. Decay of the torsional amplitude. The pendulum was given a free oscillation amplitude of $886.4\pm0.1$ $\mu$rad; 17.6 h later the amplitude had decayed to $780.6\pm0.1$ $\mu$rad, corresponding to a decay time of $5.00\times10^5$ s.

A low-mass, Au-coated ''parking platform'' was located inside the inner magnetic shield 2 cm below the pendulum. This was designed to catch the pendulum if a suspension fiber failed and to hold the pendulum during test-body interchanges.

### E. Autocollimator

The pendulum twist was monitored by an autocollimator shown schematically in Fig. 6. Light from a stabilized diode laser passed through a circular collimator, was deflected 90° by a beam splitter and formed into a 6 mm diameter parallel beam by a plano-convex lens. This beam was reflected twice by 90° and directed onto one of the Au-coated corner mirrors on the pendulum. The reflected light passed back through the lens and beam splitter and was focused onto a linear position-sensitive detector. The detector currents were amplified and combined in low-noise electronics to produce outputs $\Delta=(L-R)$ and $\Sigma=(L+R)$ where $L$ and $R$ were the currents from the two ends of the detector. The laser intensity was square-wave modulated at 146 Hz so that lock-in amplifiers could be used on the $\Delta$ and $\Sigma$ signals. The $\Delta$ and $\Sigma$ signals were digitized in 16-bit analog-to-digital coventers (ADCs) and the pendulum twist, $\theta$, was obtained from

$$\theta=h_1(\Delta/\Sigma)+h_2(\Delta/\Sigma)^2. \tag{7}$$

This system gave a twist signal that was unaffected by linear displacement of the pendulum or by its ''swing'' motion. The calibration of the $h_1$ and $h_2$ coefficients is described in Sec. III A.

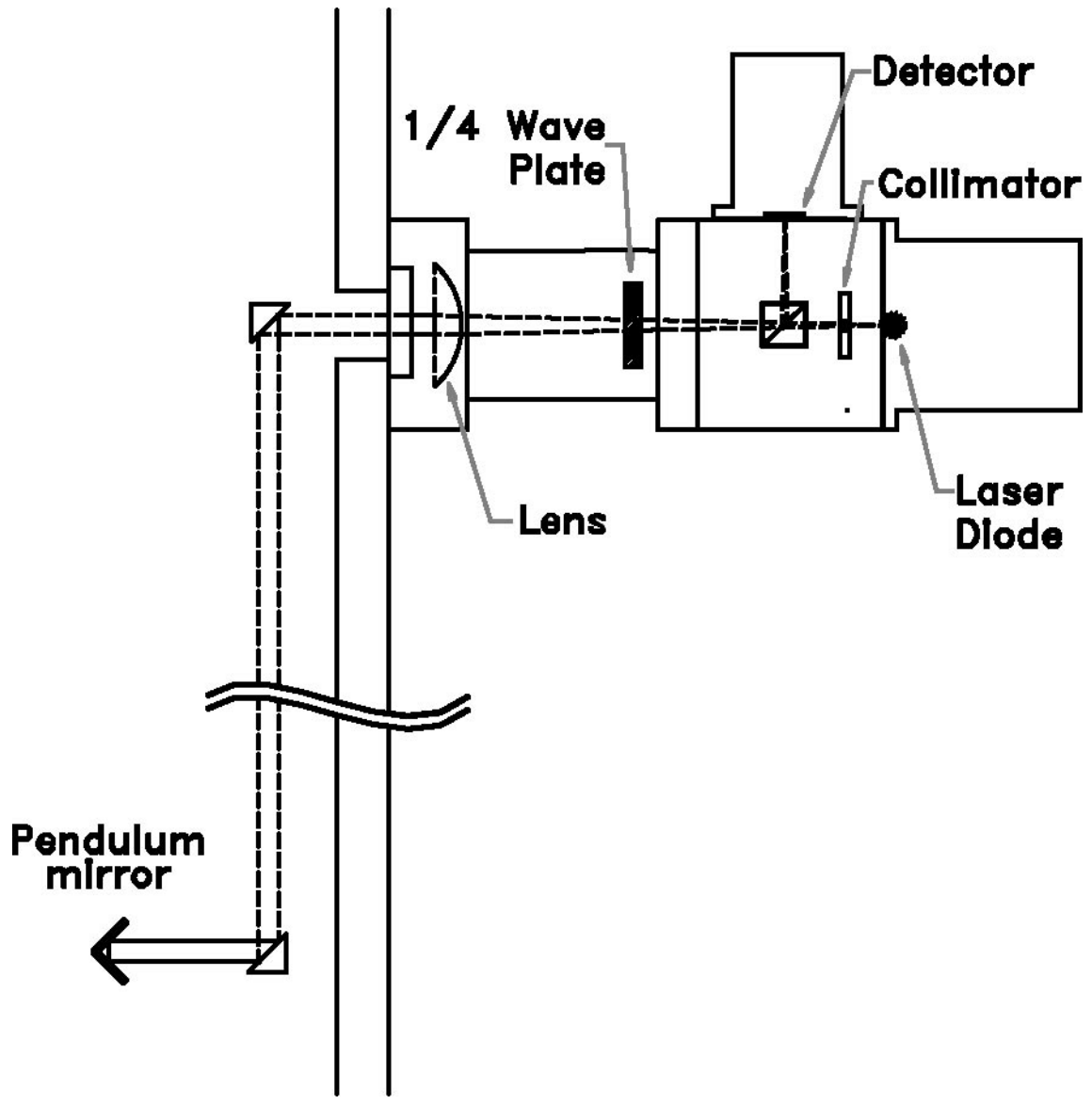

FIG. 6. Schematic diagram of the autocollimator used in the 1997–1998 data.

The laser intensity was adjusted so that 3 $\mu$W of light was received at the autocollimator detector. If the beam were mis-centered horizontally on the suspension fiber axis by an amount $\delta$, the radiation pressure torque on the pendulum would be $T_{\rm rad}=2P\delta/c$, where $P$ is the optical power hitting the mirror. The induced pendulum twist is fairly small; a misalignment $\delta=0.5$ mm would produce a twist $\theta_{\rm rad}=T_{\rm rad}/\kappa=2.9$ nrad. Because this twist was independent of test-body *configuration* and essentially independent of attractor angle [15], it did not produce a significant systematic error.

The 1996 autocollimator had a 780 nm laser [16], a dielectric beam splitter, a 5 mm long position-sensitive detector [17], and a lens with a focal length 20 cm. When this autocollimator was directed at a retro-reflector, its noise at the 0.9 mHz signal frequency was 4.5 nrad $\sqrt{\rm day}$. The 1997–1998 autocollimator used a different 780 nm laser [18], a polarization-dependent cube beam splitter, a 30 cm focal length lens, a $\lambda/4$ plate, and a 3 mm long detector [19]. When aimed at a retro-reflector this unit had a noise at our 0.9 mHz signal frequency of 0.4 nrad $\sqrt{\rm day}$.

### F. Shielding

#### 1. Electromagnetic shielding

The pendulum was surrounded by two layers of annealed mu-metal shielding. A 14 cm diameter, Au-coated, inner magnetic shield inside the vacuum vessel acted as an electrostatic shield as well. The 18 cm diameter outer shield was located just outside the vacuum vessel. Both shields were grounded in only one spot to prevent electro-chemical currents from flowing through the shields. Based on measurements with similar shields in our Eöt-Wash II balance [4], we expect that the two shields together reduced the magnetic field at the torsion pendulum by a factor of more than $10^3$.

#### 2. Thermal and acoustic shielding

The apparatus resided in a large room whose daily temperature variation was $\lesssim 1°$ C. The entire torsion balance and rotating attractor were enclosed in a double-walled wooden housing with insulating foam filling the space between the walls. This massive housing provided both thermal and acoustic shielding. In the 1996 experiment, three automobile radiators with fans were placed inside the housing and constant-temperature water from a commercial device [20] was piped through the radiators. In the 1997–1998 experiment, the heat exchangers and fans were moved outside the housing and the temperature-controlled air was fed into and out of the housing in insulated ducts. Both systems had similar performance and limited the daily air-temperature variations inside the housing to $\pm 0.01$ K.

An inner layer of active thermal shielding completely surrounded the torsion balance itself. This layer consisted of a copper shield with constant-temperature water from a second constant-temperature bath circulating along it. The outer surface of this shield was covered with 1 cm of insulating foam except in the region between the torsion balance and the uranium attractor (where space was at a premium) where the insulation thickness was only 2 mm. In addition a 0.8 mm thick copper annulus was placed on the turntable axis just inside the uranium attractor to homogenize any azimuthal temperature variations in the attractor. The temperature variations, at the attractor rotation frequency, of the inner shield and of the air inside the shield were 0.53 mK and 0.08 mK, respectively. A shiny Ni-coated Cu tube surrounding the torsion fiber inside the vacuum vessel provided substantial passive shielding against short- and-moderate-term temperature variations of the fiber. This tube is omitted from Fig. 1 for clarity.

### G. Data monitoring

A personal computer with a 16-bit data acquisition board [21] monitored the pendulum twist, the attractor angle, the tilts of the torsion balance and of the attractor turntable, and 13 temperatures. Four temperature sensors were mounted on the vacuum vessel, 7 measured air temperatures inside and outside of the thermal shields, and 2 monitored the constant-temperature water supplies. Horizontal and vertical seismic accelerations, and the pressure in the vacuum vessel, were monitored in some of the data. All these quantities were periodically digitized and recorded for off-line analysis.

## III. EXPERIMENTAL PROTOCOLS AND ANALYSIS

### A. Calibrations

#### 1. Pendulum twist, period, and damping time

The pendulum angular deflection scale was calibrated in 3 independent ways: by rotating the upper fiber attachment point through a known angle, by displacing the autocollimator's position-sensitive detector by a known amount, and by applying known gravitational torques to the pendulum.

An abrupt rotation of the upper fiber attachment (typically by 500 $\mu$rad) excited torsional oscillations of the pendulum as shown in Fig. 7. A fit of this dynamic-calibration data to a damped sine wave yielded precise values for the free-oscillation period, $\tau_0$, and amplitude damping time, $\tau_{\rm d}$. The fit residuals revealed any non-linearities of the angular deflection scale and were used to fix $h_2/h_1$, the ratio of calibration coefficients in Eq. (7). However, the absolute pendulum twist calibration scattered by about $\pm 5\%$ because of play in the rotation stage.

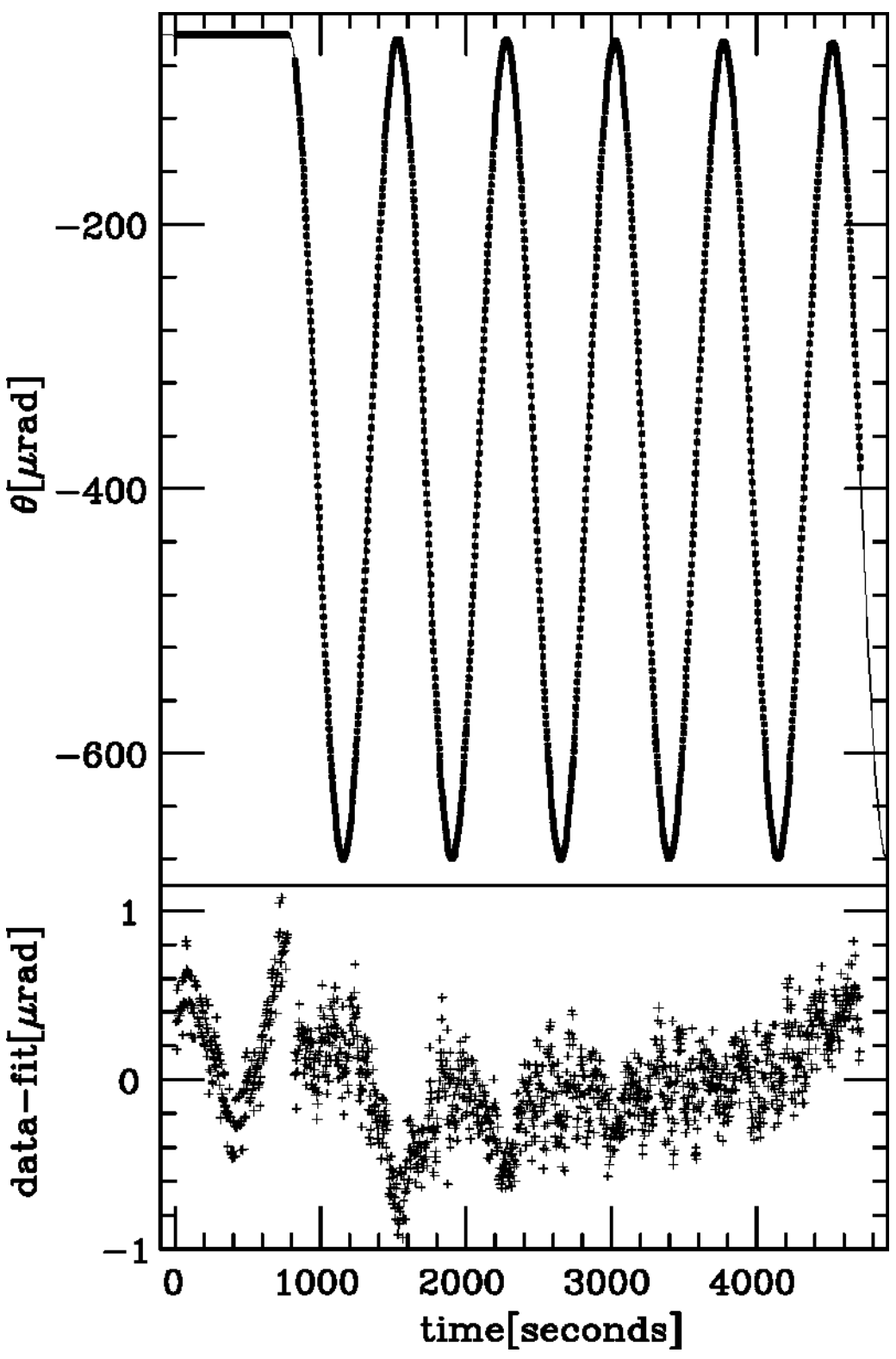

FIG. 7. Dynamic calibration data. At $t=770$ s the rotation stage was twisted by 319 $\mu$rad. The resulting free torsional oscillation (the small squares in the upper panel) was fitted by a damped oscillation (the smooth curve). The expanded scale in the lower panel shows the fit residuals.

Absolute calibrations of the angular deflection scale were made by reflecting the autocollimator beam from a corner cube and observing the $\theta$ signal as the autocollimator detector was translated along its position-sensitive ($x$) axis in 0.10 mm steps. We then varied the $h_1$ and $h_2$ coefficients in Eq. (7) to obtain the best fit of the $\theta$ points to the corresponding calibration values, $\theta(x)=2\tan^{-1}x/r$, where $r$ is the distance between the lens and detector (approximately a focal length). The $h_2/h_1$ ratio extracted from this fit agreed well with the ratio extracted from the dynamic calibrations.

Direct gravitational calibrations of the torque scale were performed periodically. For these calibrations we created known $Q_{44}$ fields by adding 4 identical masses, equally spaced azimuthally, to the attractor's midplane. The predicted torque from the coupling of the $Q_{44}$ field to the calculated $q_{44}$ moment was within 0.5% of the measured value. Calibrations were made with $Q_{44}$ gradients of several different magnitudes, and so provided a check of the linearity of the torque scale. The results are shown in Fig. 8.

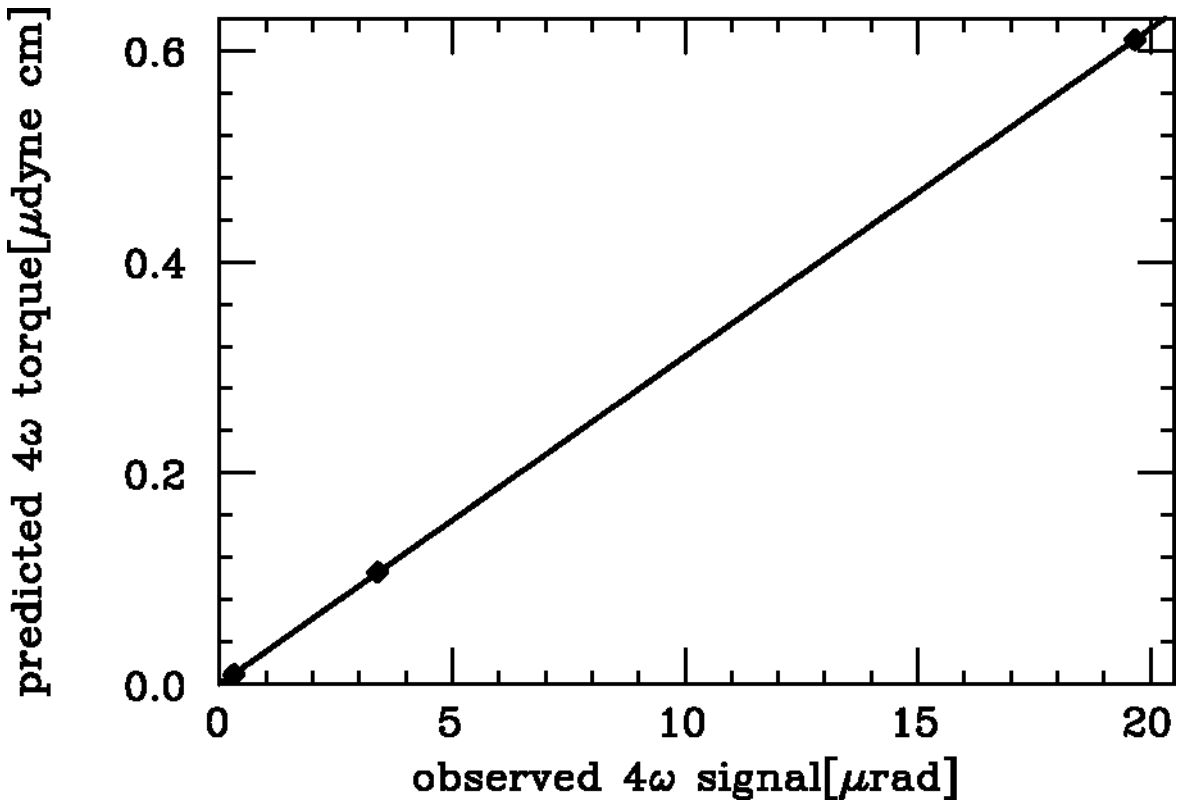

FIG. 8. Test of the linearity of the torque scale, showing the $4\omega$ signal from direct gravitational calibrations using a series of successively weaker $Q_{44}$ attractors. The errors are smaller than the size of the points.

### *2. Fiber torsion constant*

Two similar suspension fibers were used in this work. After a small earthquake broke fiber 1 halfway through the 1996 data set it was replaced by fiber 2, which was used for the remainder of the 1996 and all of the 1997–1998 data. The fiber torsion constants were inferred from the measured $\tau_0$'s and the calculated moment of inertia of the pendulum which was known to better than 0.7%. The torsion constants of fibers 1 and 2 were $\kappa=0.0313\pm0.0001$ and $0.0311\pm0.0001$ erg/rad, respectively; the errors are statistical and do not include the 0.7% scale factor uncertainty from the moment of inertia.

### *3. Temperatures*

Temperatures were measured with solid-state devices [22] whose currents were proportional to the absolute temperature. Stable currents equivalent to a temperature of about 293 K were subtracted from the individual sensor outputs. The sensors were calibrated at two temperatures by mounting them on a temperature-controlled Cu block.

### *4. Tilts*

The torsion balance tilt and the flex of one of the turntable legs were measured with a pair of 2-axis electronic levels [23]. As we were sensitive only to changes in the tilts, calibration of the electronic levels was straightforward; we mounted them on platforms having leveling screws with a known lever arm and displaced the screws by known amounts.

## B. Procedure

### *1. Setup and alignment*

The pendulum was centered on the rotating attractor by gravitational means. A horizontal displacement of the pendulum from the attractor center, $\vec{\epsilon}$, produced a $5\omega$ torque (see Refs. [24,25])

TABLE I. Configuration-independent gravitational multipole moments of the normal pendulum. The average values are obtained from measurements after each *configuration* change; the errors are typical statistical uncertainties of an individual measurement. The 1997–1998 stray moments are generally smaller than the 1996 moments because of better adjustment of the trim screws and reduced pendulum tip.

| Moment | Design value | 1996 average value | 1997–1998 average value |
|---|---|---|---|
| $\lvert q_{00}\rvert$ | 17.971 g | $17.971\pm0.003$ g | Same |
| $\lvert q_{10}\rvert$ | $\approx 0.0$ | | |
| $\lvert q_{11}\rvert$ | $\approx 0.0$ | | |
| $\lvert q_{20}\rvert$ | $\approx 0.0$ | | |
| $\lvert q_{21}\rvert$ | $\approx 0.0$ | $0.0622\pm0.0002$ g cm$^2$ | $0.0331\pm0.0001$ g cm$^2$ |
| $\lvert q_{22}\rvert$ | $\approx 0.0$ | $0.009\pm0.002$ g cm$^2$ | |
| $\lvert q_{30}\rvert$ | $\approx 0.0$ | | |
| $\lvert q_{31}\rvert$ | $\approx 0.0$ | $0.197\pm0.002$ g cm$^3$ | $0.196\pm0.002$ g cm$^3$ |
| $\lvert q_{32}\rvert$ | $\approx 0.0$ | | |
| $\lvert q_{33}\rvert$ | $\approx 0.0$ | $0.291\pm.001$ g cm$^3$ | $0.181\pm0.002$ g cm$^3$ |
| $\lvert q_{40}\rvert$ | 4590 g cm$^4$ | | |
| $\lvert q_{41}\rvert$ | $\approx 0.0$ | $56\pm2$ g cm$^4$ | $22\pm2$ g cm$^4$ |
| $\lvert q_{42}\rvert$ | $\approx 0.0$ | | |
| $\lvert q_{43}\rvert$ | $\approx 0.0$ | $13.1\pm0.5$ g cm$^4$ | $5.1\pm0.5$ g cm$^4$ |
| $\lvert q_{44}\rvert$ | $-1695$ g cm$^4$ | | |

$$T_{\mathrm{g}}^{m=5}=-40\pi G\sqrt{\frac{5}{22}}[\mathrm{Re}(\{\epsilon_x+i\epsilon_y\}q_{44}Q_{55})\sin 5\phi + \mathrm{Im}(\{\epsilon_x+i\epsilon_y\}q_{44}Q_{55})\cos 5\phi], \quad (8)$$

where $q_{44}$ and $Q_{55}$, given in Tables I and II respectively, are large designed-in moments. Figure 9 shows how this $5\omega$ signal was used to align the pendulum to within $\pm 40$ $\mu$m of its proper horizontal position. The pendulum was centered vertically by installing $q_{21}$ gradiometer bodies (see Fig. 10) on the pendulum and putting the attractor in the $Q_{31}$ *state* (see Fig. 4). In these circumstances an upward vertical displacement of the pendulum from the attractor center, $\eta$, produced a $1\omega$ torque [24]

TABLE II. Gravitational moments of the attractor in its normal *state*. The $Q_{21}$, $Q_{31}$, and $Q_{41}$ stray moments varied slightly from one *state* change to another because the attractor *states* were not precisely reproducible. The measured moments are averages over the *state* changes; the errors are typical statistical uncertainties in an individual measurement.

| Moment | Design value | 1996 measured value | 1997–1998 measured value |
|---|---|---|---|
| $Q_{00}$ | $2.52\times10^4$ g/cm | $2.52\times10^4$ g/cm | |
| $Q_{10}$ | $\approx 0.0$ | | |
| $Q_{11}$ | $-481$ g/cm$^2$ | | |
| $Q_{20}$ | $-9.4$ g/cm$^3$ | | |
| $Q_{21}$ | $\approx 0.0$ | $(9.9\pm0.1)\times10^{-3}$ g/cm$^3$ | $(1.42\pm0.03)\times10^{-3}$ g/cm$^3$ |
| $Q_{22}$ | $\approx 0.0$ | $(0.11\pm0.02)$ g/cm$^3$ | |
| $Q_{30}$ | $\approx 0.0$ | | |
| $Q_{31}$ | $\approx 0.0$ | $(1.1\pm0.1)\times10^{-3}$ g/cm$^4$ | $(1.20\pm0.09)\times10^{-3}$ g/cm$^4$ |
| $Q_{32}$ | $\approx 0.0$ | | |
| $Q_{33}$ | 0.274 g/cm$^4$ | | |
| $Q_{40}$ | $-0.007$ g/cm$^5$ | | |
| $Q_{41}$ | $\approx 0.0$ | $4.1\times10^{-5}$ g/cm$^5$ | $(4.69\pm0.34)\times10^{-5}$ g/cm$^5$ |
| $Q_{42}$ | $\approx 0.0$ | | |
| $Q_{43}$ | $\approx 0.0$ | | |
| $Q_{44}$ | $\approx 0.0$ | $(3.9\pm0.2)\times10^{-6}$ g/cm$^5$ | $(12.4\pm0.1)\times10^{-6}$ g/cm$^5$ [a] |
| $Q_{51}$ | $-6.5\times10^{-6}$ g/cm$^6$ | | |
| $Q_{55}$ | $-5.3\times10^{-4}$ g/cm$^6$ | | |

[a]A deliberate weak $Q_{44}$ mass distribution was added to the attractor.

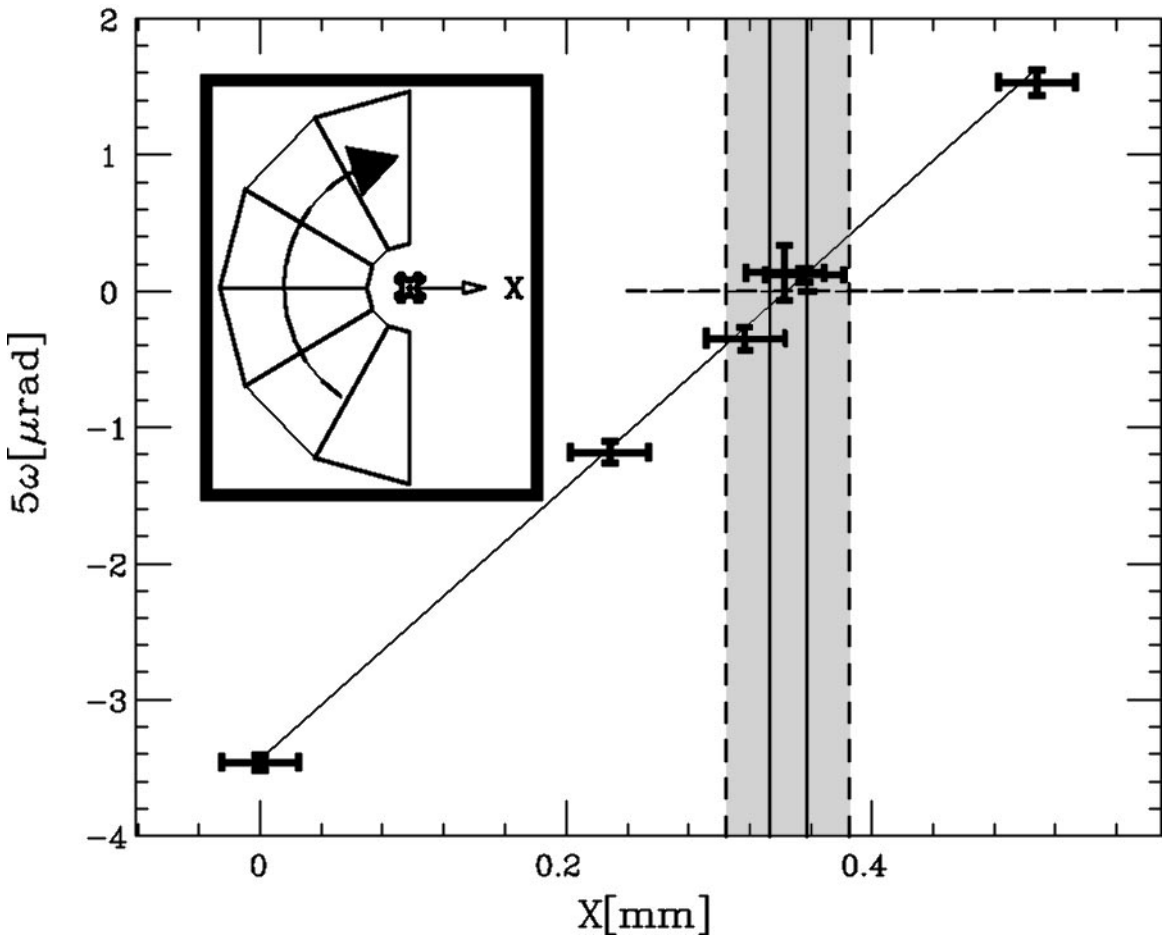

FIG. 9. Gravitational centering of the pendulum in the horizontal plane. The $5\omega$ signal was proportional to the horizontal displacement between the pendulum and the attractor rotation axis. The data points were fitted with a straight line whose slope was fixed at the value predicted by the calculated $q_{44}$ and $Q_{55}$ moments of the pendulum and attractor, respectively. The uncertainty in the *X* intercept, $\pm 12$ $\mu$m, is shown by the vertical lines. The shaded band shows the $\pm 40$ $\mu$m total centering uncertainty taking into account the variation of the pendulum's position as the upper fiber attachment was rotated to different *angles.*

$$T_{\rm g}^{m=1} = 16\pi G\sqrt{\frac{2}{35}}\eta[\mathrm{Re}(q_{21}Q_{31})\sin\phi + \mathrm{Im}(q_{21}Q_{31})\cos\phi], \tag{9}$$

where $q_{21}$ and $Q_{31}$ are given in Tables III and IV, respectively. As shown in Fig. 11, we could place the pendulum to within $\pm 10$ $\mu$m of the vertical center of the $Q_{31}$ attractor by adjusting the pendulum height until the induced $1\omega$ torque vanished. However, the vertical centering with respect to the attractor's normal *state* was only good to $\pm 60$ $\mu$m.

### *2. Data-taking protocol*

Before taking data, the fiber unwinding rate was reduced to $\leqslant 1$ $\mu$rad/h by heating the entire apparatus to $\sim 50°$ C

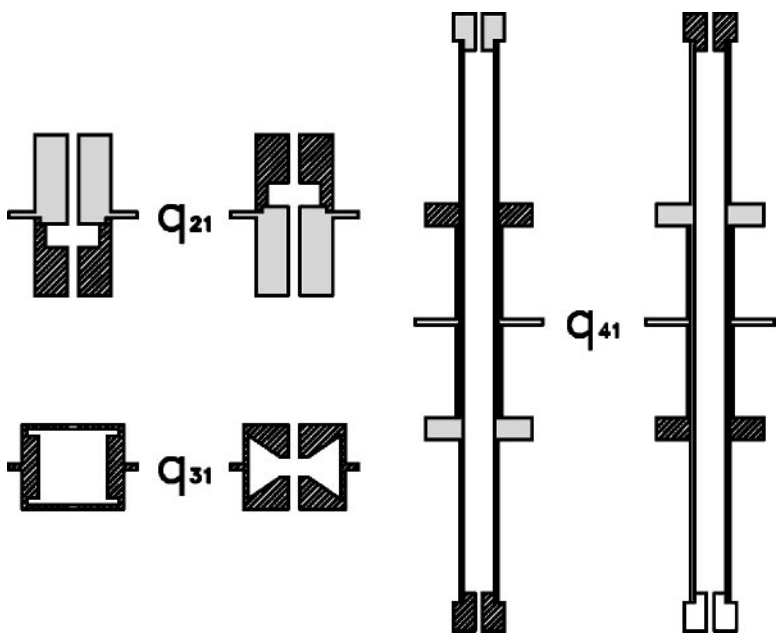

FIG. 10. Cross-sectional views of the cylindrically symmetric bodies used to turn the pendulum into a $q_{21}$, $q_{31}$, or $q_{41}$ gradiometer. The hatching denotes Cu and the light shading Al. The gradiometer pendulums had vanishing composition dipole moments so that they were insensitive to composition-dependent interactions.

TABLE III. Gravitational multipole moments of the gradiometer pendulums.

| Gradiometer | Moment | Design value |
|---|---|---|
| ''$q_{21}$'' | $q_{21}$ | 21.2 g cm$^2$ |
| | $q_{41}$ | $-355$ g cm$^4$ |
| ''$q_{31}$'' | $q_{31}$ | $-5.51$ g cm$^3$ |
| | $q_{51}$ | 154 g cm$^5$ |
| ''$q_{41}$'' [a] | $q_{21}$ | 0.171 g cm$^2$ |
| | $q_{41}$ | $-987$ g cm$^4$ |

[a]Gradiometer pendulum used in the 1997–1998 data.

for 24 h while the pendulum was freely suspended in vacuum. This procedure had to be repeated whenever the stress on the fiber was released by ''parking'' the pendulum; reapplying the stress to the fiber invariably caused an increased unwinding rate that had to be annealed out. The unwinding rate was typically 0.6 $\mu$rad/h (0.2 $\mu$rad/h) in the 1996 (1997–1998) experiment,

Essentially equal amounts of data were taken for two opposite *configurations*, $\mathcal{A}$ and $\mathcal{B}$, of the test-body composition dipole on the pendulum tray, and for two diametrically opposed *angles*, $\mathcal{N}$ and $\mathcal{R}$, of the pendulum tray with respect to the autocollimator (see Fig. 12). We followed a fixed protocol to ensure that the sensitivities to systematic errors were always measured without disturbing the pendulum *configurations* or attractor *states* used in acquiring UFF data.

First, we measured the residual magnetic moment of the pendulum in one *configuration*, say $\mathcal{A}$. Then we used the $Q_{21}$, $\tilde{Q}_{21}$, and $Q_{31}$ attractor *states* (shown in Fig. 4) to measure the pendulum's $q_{21}$, $q_{31}$, and $q_{43}$ moments. (The $q_{43}$ moment was used to infer the pendulum ''tip'' angle as described in Sec. III E.) We then returned the attractor to its normal *state* and took UFF data with pendulum *angles* $\mathcal{N}$ and $\mathcal{R}$. Then we opened the vacuum vessel to install the $q_{21}$ gradiometer test bodies (the gradiometer bodies are shown in

TABLE IV. Gravitational moments of the attractor in its $Q_{21}$, $\tilde{Q}_{21}$, and $Q_{31}$ *states.*

| State | Moment | Design value |
|---|---|---|
| $Q_{21}$ | $Q_{21}$ | 16.3 g/cm$^3$ |
| | $Q_{31}$ | 0.252 g/cm$^4$ |
| | $Q_{41}$ | $-0.011$ g/cm$^5$ |
| | $Q_{51}$ | $\approx 0$ |
| $\tilde{Q}_{21}$ | $Q_{21}$ | 16.3 g/cm$^3$ |
| | $Q_{31}$ | $-0.252$ g/cm$^4$ |
| | $Q_{41}$ | $-0.011$ g/cm$^5$ |
| | $Q_{51}$ | $\approx 0$ |
| $Q_{31}$ | $Q_{21}$ | $\approx 0$ |
| | $Q_{31}$ | $-0.501$ g/cm$^4$ |
| | $Q_{41}$ | $\approx 0$ |
| | $Q_{51}$ | 0.0021 g/cm$^6$ |

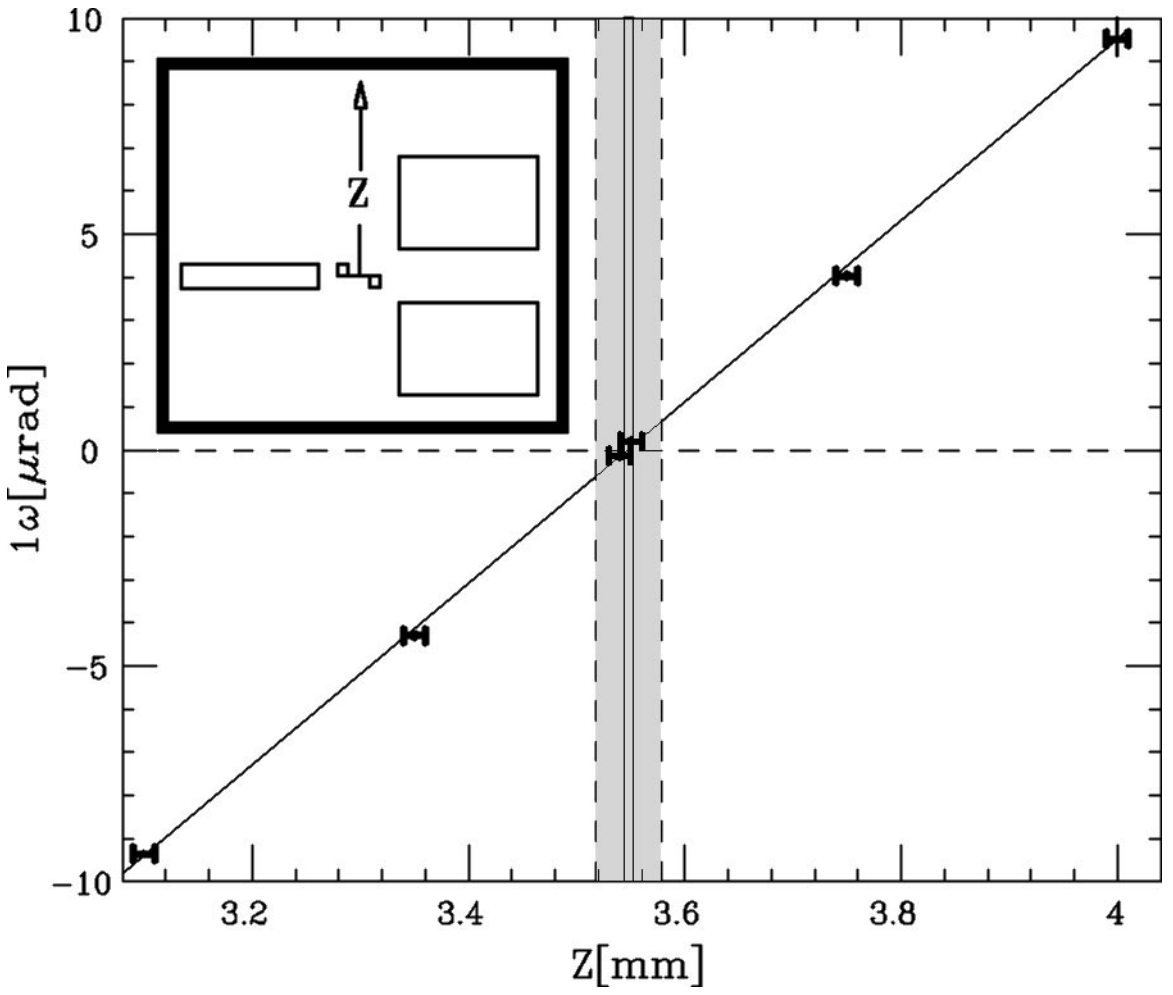

FIG. 11. Gravitational centering of the pendulum along the vertical axis of the attractor. The pendulum was configured as a $q_{21}$ gradiometer and the attractor was placed in the $Q_{31}$ *state*. Under these conditions the $1\omega$ signal was proportional to the vertical displacement between the pendulum and the symmetry plane of the attractor. The data points were fitted with a straight line whose slope was fixed at the value predicted by the calculated $q_{21}$ moment of the pendulum and the calculated $Q_{31}$ gradient of the attractor. The uncertainty in the intercept is shown by the vertical lines. The $\pm 60$ $\mu$m shaded region shows the realistic vertical centering uncertainty which takes into account imperfections in rotating the attractor between its $Q_{31}$ and normal *states*.

Fig. 10) and measured the stray $Q_{21}$ field of the attractor. We repeated this process with $q_{31}$ and $q_{41}$ gradiometer test bodies to measure the attractor's stray $Q_{31}$ and $Q_{41}$ fields. (We made only a single measurement of the $Q_{41}$ field in the 1996 experiment, using the $q_{41}$ gradiometers shown in Fig. 6 of Ref. [4].) We then removed the torsion balance and measured the unshielded magnetic field at the usual site of the pendulum. Next, we reinstalled the torsion balance with the test bodies in the opposite *configuration*, say $\mathcal{B}$, and took UFF data at *angles* $\mathcal{N}$ and $\mathcal{R}$. Then we changed attractor *states* and measured the pendulum's gravitational moments in the second *configuration*. Finally, we measured the magnetic moment of the pendulum in the second *configuration* and repeated the cycle. This protocol ensured that we measured the gravitational and magnetic moments and fields of the pendulum and attractor actually present in the UFF data.

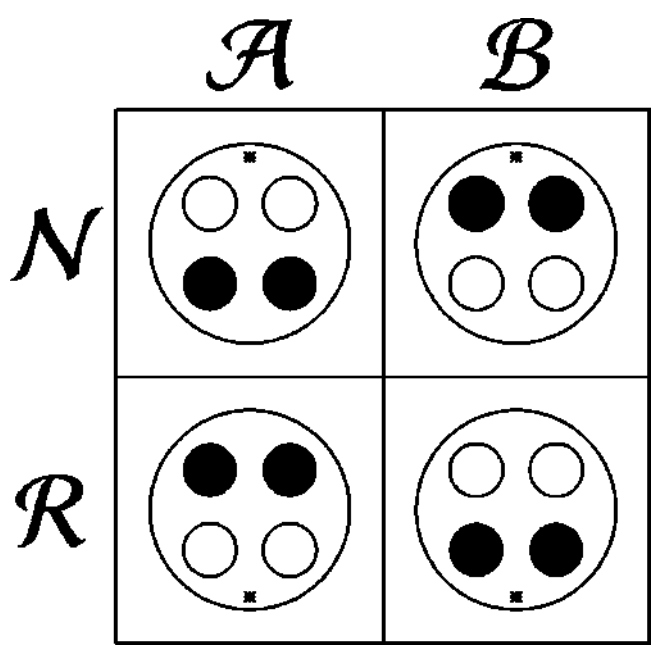

FIG. 12. Schematic drawing of the pendulum *configurations* and *angles*. The dot refers to a fixed point on the pendulum tray.

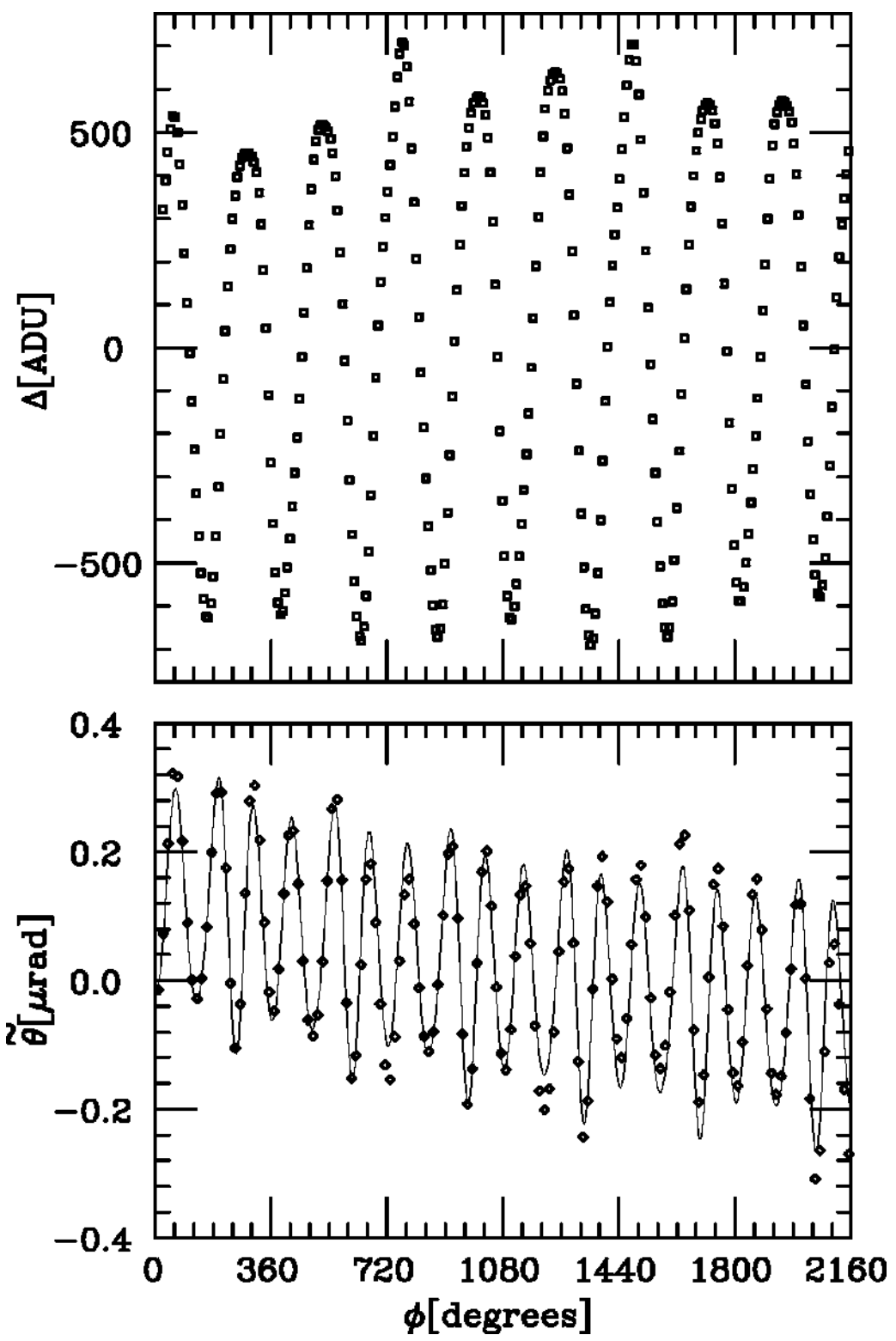

FIG. 13. Pendulum twist data from a typical cut. Upper panel: autocollimator output, $\Delta$, vs attractor angle, $\phi$. The data are dominated by the free torsional oscillation. Lower panel: the filtered and averaged twist $\tilde{\theta}$ along with the harmonic fit which yielded amplitudes of $1.8\pm5.5$, $25.3\pm5.5$, $163.5\pm5.5$, $37.2\pm5.5$, and $17.5 \pm5.5$ nrad for the $1\omega$, $2\omega$, $3\omega$, $4\omega$, and $5\omega$ harmonic signals. (The steady drift has opposite signs in the 2 panels because of our sign convention for converting $\Delta$ into $\tilde{\theta}$.)

### C. Data analysis

#### 1. Harmonic analysis of the pendulum twist

The attractor revolution period and data-recording interval were set to $\tau_A = 1.5\tau_0 \approx 1121$ s and $\tau_R = \tau_A/48 \approx 23$ s, respectively, where $\tau_0$ is the period of free torsional oscillations. Data were accumulated continuously, and later cut into segments containing exactly 6 revolutions of the attractor and therefore 288 data points. Figure 13 shows twist data for a typical "cut." The twist signals $\theta(t)$ from each cut were analyzed by first applying a digital filter that removed the free torsional oscillations,

$$\theta_f(t) = [\theta(t-\tau_0/4) + \theta(t+\tau_0/4)]/2, \tag{10}$$

and then averaging the $\theta_f$ values of two successive points to obtain a filtered, averaged twist, $\tilde{\theta}$, also shown in Fig. 13. We fitted $\tilde{\theta}$ as a function of $\phi$ (recall that $\phi = \omega t$ is the angle between the composition dipole and the center of the uranium attractor):

TABLE V. Attenuation and phase shift of harmonic twist signals from pendulum inertia, electronic time constants, and digital filtering + averaging. These values correspond to an electronic time constant of 26.2 s, a data sampling interval of 23.3 s, $\tau_0=747.4$ s and $\tau_d=4.9\times10^5$ s. Phase shifts are given in turntable degrees.

| $n$ | Pendulum inertia | | Time constant | | Digital filter + averaging | | Total | |
|---|---|---|---|---|---|---|---|---|
| | Atten. | Phase shift | Atten. | Phase shift | Atten. | Phase shift | Atten. | Phase shift |
| 1 | 1.793 | −0.1° | 0.979 | −16.7° | 0.499 | 0° | 0.876 | −16.8° |
| 2 | 1.301 | −89.9° | 0.921 | −16.3° | 0.496 | 90° | 0.594 | −16.2° |
| 3 | 0.336 | −60.0° | 0.838 | −15.8° | 0.981 | 60° | 0.276 | −15.8° |
| 4 | 0.165 | −45.0° | 0.744 | −15.2° | 0.483 | 45° | 0.059 | −15.2° |
| 5 | 0.099 | −36.0° | 0.651 | −14.5° | 0.473 | 0° | 0.031 | −50.5° |

$$\tilde{\theta}(\phi)=\sum_{n=1}^{5}[a_n\sin n\phi+b_n\cos n\phi]+\sum_{m=1}^{3}d_mP_m(x), \tag{11}$$

where the Legendre polynomials, $P_m$, whose arguments, $x$, are proportional to time, accounted for the continuous unwinding of the torsion fiber apparent in Fig. 13. The harmonic sum in Eq. (11) extends to $n=5$ because a $5\omega$ torque was produced when the 4-fold symmetric pendulum was misaligned with respect to the revolution axis of the attractor (see Sec. III B 1 above). Finally, the $a_n$ and $b_n$ coefficients in Eq. (11) were corrected for pendulum inertia, electronic time constants, digital filtering and averaging, and for gravity gradient torques, yielding corrected coefficients $\tilde{a}_n$ and $\tilde{b}_n$ used in all subsequent calculations.

#### 2. Phase shift and attenuation corrections

The corrections for phase shift and attenuation arising from pendulum inertia, the two-pole filters of the amplifiers, and digital filtering and averaging were straightforward. The inertia correction depended upon $\tau_0$ and $\tau_{\rm d}$, which were known precisely from the dynamic calibrations. The electronic attenuation and phase shift were functions of the amplifier time constants that we measured by applying step inputs to the amplifiers and analyzing their responses. In addition, we accounted for phase shift and attenuation from the digital filter and the averaging process. Characteristic values for the attenuation and phase shift corrections are listed in Table V.

The phase-shift corrections were tested by comparing data taken with both senses of attractor rotation, which gave phase shifts of opposite sign. The largest signal in these runs had raw phase shifts of $\approx+109°$ and $\approx-109°$. After the phase-shift correction the signal phases agreed within the 0.7° measurement error. The attenuation corrections were verified by the direct gravitational torque measurements, discussed in Sec III A 1, which agreed to within 0.5% with calibrations based on the measured $\tau_0$, the calculated pendulum rotational inertia and $q_{44}$ moment, the known masses and positions of the Pb blocks of the $Q_{44}$ attractor, and the measured autocollimator angular deflection scale.

#### 3. Gravity-gradient corrections

Gravity gradient corrections were computed from Eq. (6) using the measured $Q_{21},Q_{31}$ and $Q_{41}$ fields of the normal attractor and the measured $q_{21}$ and $q_{31}$ moments of the normal pendulum. It was not practical to measure the pendulum's $q_{41}$ moments directly, so these were inferred from its $q_{21}$ and $q_{31}$ moments as discussed below. Because the $Q_{41}$ fields in the 1996 experiment were measured only once and not for each *configuration* change, we could not correct the 1996 data for the $l=4$ gradients; instead we added the entire $l=4$ torque from that single measurement to the systematic error budget.

The gravity-gradient corrections to our UFF-violating signal depend only on the *configuration*-dependent gravity-gradient torque, i.e. on the changes in the gravity-gradient torque associated with changes in the test-body *configurations*. The pendulum's *configuration*-dependent $q_{41}$ moment could be inferred from the measured *configuration*-dependent $q_{21}$ and $q_{31}$ moments summarized in Table VI because all these moments arose primarily from test-body misalignments rather than from small-scale inhomogeneities. For example, a vertical test body misalignment would produce a $q_{21}$ moment as well as a $q_{41}$ moment, while a horizontal misalignment would produce a $q_{31}$ moment and also tip the pendulum slightly, generating a $q_{41}$ moment from the large $q_{40}$ value. The *configuration*-dependent $q_{21}$ moment in the 1996 fiber-1 data corresponded to one test body misaligned vertically by 14 $\mu$m, while the *configuration*-dependent $q_{31}$ moment corresponded to one test body misaligned radially by 18 $\mu$m. This radial misalignment changed the pendulum tip by 56 $\mu$rad and rotated the as-designed $q_{40}$ moment into a $q_{41}$ moment. We used our mul-tipole analysis program, which contains a complete model of the pendulum and attractor, to infer the stray *configuration*-dependent $q_{41}$ moment. The net $l=4$ contributions to our 1996 and 1997–1998 signals were 0.18 nrad and $+0.15\pm0.02$ nrad, respectively. The small size of the $l=4$ gravitational torques justified our neglect of $l>4$ couplings; gravity gradient torques scale as $(r/R)^L$, where $r\sim3$ cm and $R\sim27$ cm are characteristic dimensions of the pendulum and attractor, respectively.

TABLE VI. Net *configuration*-dependent gravitational multipole moments of the normal pendulum. These are defined as $\Delta q_{lm}=|q_{lm}^{\mathcal{A}}-q_{lm}^{\mathcal{B}}|/2$.

| Moment | 1996 value | 1997–1998 value |
|---|---|---|
| $\Delta q_{21}$ | $0.033773\pm0.00006$ g cm$^2$ | $0.01697\pm0.00002$ g cm$^2$ |
| $\Delta q_{31}$ | $0.124\pm0.002$ g cm$^3$ | $0.149\pm0.001$ g cm$^3$ |
| $\Delta q_{41}$[a] | $0.369\pm0.006$ g cm$^4$ | $0.370\pm0.003$ g cm$^4$ |

[a]Inferred from $\Delta q_{21}$ and $\Delta q_{31}$.

#### 4. Signal extraction

The combination of the pendulum's *configuration* and *angle* defined two *orientations*, $\oplus$ and $\ominus$, of the composition dipole in the laboratory frame. The corrected coefficients, $\tilde{a}_n$ and $\tilde{b}_n$, from $N$ pairs of ''cuts'' with opposite *orientations* of the composition dipole were combined with equal weights to generate our UFF-violating signal, $S$, and a quadrature null, $P$,

$$S=\sum_{i=1}^{N} s_i/N \qquad s_i=[(\tilde{a}_1)_i^{\oplus}+(\tilde{a}_1)_i^{\ominus}]/2$$

$$P=\sum_{i=1}^{N} p_i/N \qquad p_i=[(\tilde{b}_1)_i^{\oplus}+(\tilde{b}_1)_i^{\ominus}]/2, \tag{12}$$

in which subtractions inherent in the *configuration* and, to a lesser degree, *angle* reversals suppressed all systematic effects not explicitly associated with the the test bodies themselves. $S$ and $P$ have statistical uncertainties given by

$$\delta S=\sqrt{\sum_{i=1}^{N}[s_i-S]^2/(N-1)}$$

$$\delta P=\sqrt{\sum_{i=1}^{N}[p_i-P]^2/(N-1)}. \tag{13}$$

### D. 1996 data

The data consist of 186 cuts taken with fiber 1 and 320 cuts taken with fiber 2, where the fiber-1 (fiber-2) data had 3 (5) changes in the *orientation* of the composition dipole. The $s_i$ and $p_i$ values from 10 data sets containing 253 pairs of ''cuts'' with opposite *orientations* of the composition dipole were were combined, using inverse error-squared weighting, to yield

$$S=-0.43\pm0.76 \text{ nrad}$$

$$P=+1.01\pm0.73 \text{ nrad}. \tag{14}$$

We verified that the statistical uncertainties in $S$ and $P$ corresponded to 68.3% confidence limits by binning all 253 $s_i$ and $p_i$ values and fitting the resulting histograms with Gaussians as shown in Fig. 14. The Gaussian fits to the $S$ and $P$ histograms were excellent, with $\chi^2$ probabilities of 88% and 90%, respectively. The the best-fit $S$ and $P$ values,

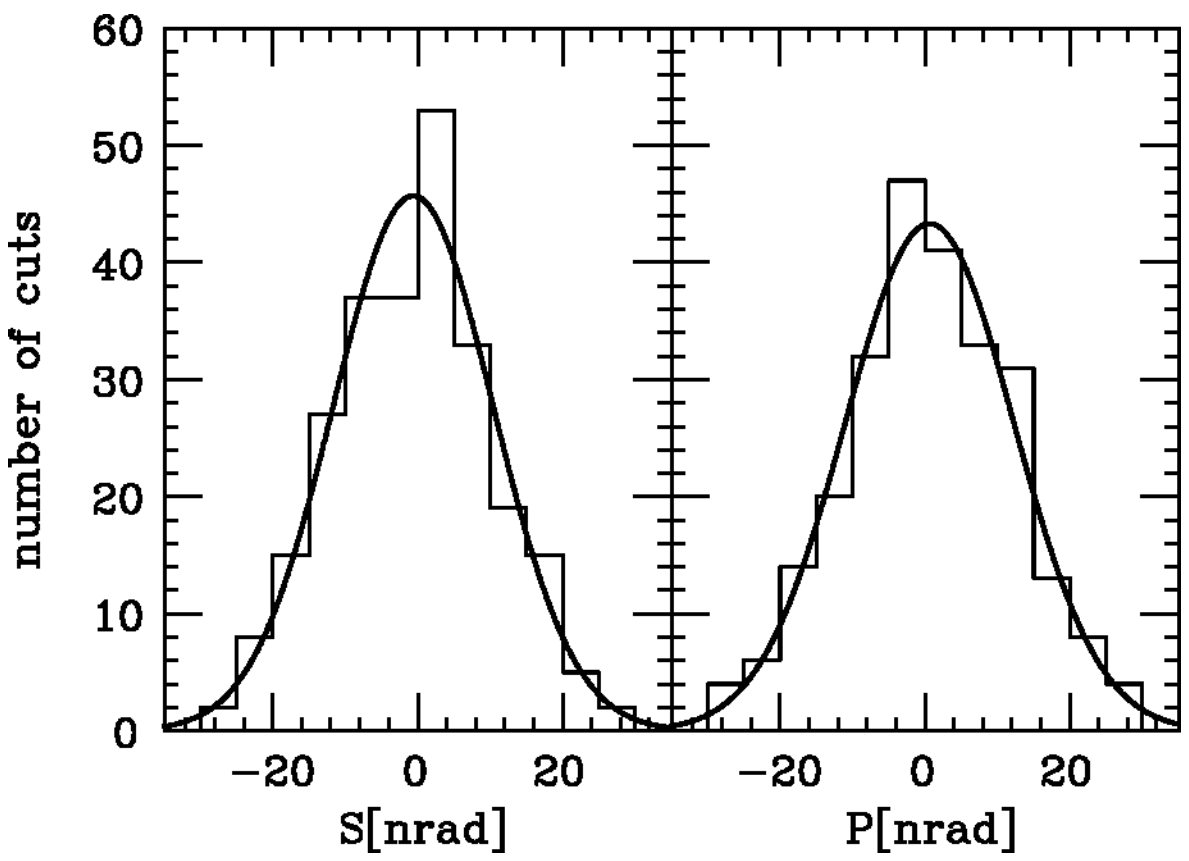

FIG. 14. 1996 results for the signal, $S$, and quadrature, $P$, values for 253 pairs of data points with opposite *orientations* of the composition dipole on the pendulum tray. The smooth curves are Gaussian fits to the histograms.

$$\langle S\rangle=-0.63\pm0.70 \text{ nrad}$$

$$\langle P\rangle=+0.64\pm0.74 \text{ nrad}, \tag{15}$$

agree well with those given in Eqs. (14). The central values and uncertainties in Eqs. (15) differ slightly from those in Eqs. (14) because in the former all 253 data pairs were weighted equally and in the latter they were not.

### E. 1997–1998 data

Several changes were made to the instrument after the 1996 data were taken.

(1) The test body masses were made more nearly equal by evaporating additional gold onto the Pb bodies. After this adjustment the Pb and Cu body masses were identical to $\pm27$ $\mu$g; in the 1996 data the Cu body masses exceeded those of the Pb bodies by $\sim150$ $\mu$g.

(2) The fiber attachment to the pendulum was realigned to reduce the pendulum's tip angle. The tip angle, $\beta=5\pm1$ mrad, in the 1996 experiment was measured by observing the hanging pendulum with a theodolite. This tip angle agreed with the more precise value, $\beta=\sin^{-1}(q_{43}/\sqrt{2}q_{44})=5.13\pm0.01$ mrad, obtained by assuming that the pendulum's $q_{43}$ moment arose from a rotation of its large, calculated $q_{44}$ moment. The tip in the 1997–1998 experiment, deduced from $q_{43}$, was reduced to $2.07\pm0.01$ mrad.

(3) The autocollimator was upgraded as described in Sec. II E.

(4) The reproduceability of the attractor was improved. In the 1996 data, the attractor's stray moments were often larger than the original ''tuned out'' values; the increased moments were traced to small changes in the attractor geometry caused by rotating its segments between the normal and the $Q_{21}$ and $Q_{31}$ *states*. We addressed this problem by installing a better dial-gauge system for monitoring the relative azimuthal angles of the three attractor segments, and by improving the mechanism for rotating the attractor segments. With this system we could ensure that the attractor tilt and azimuthal alignment returned to their nominal values within

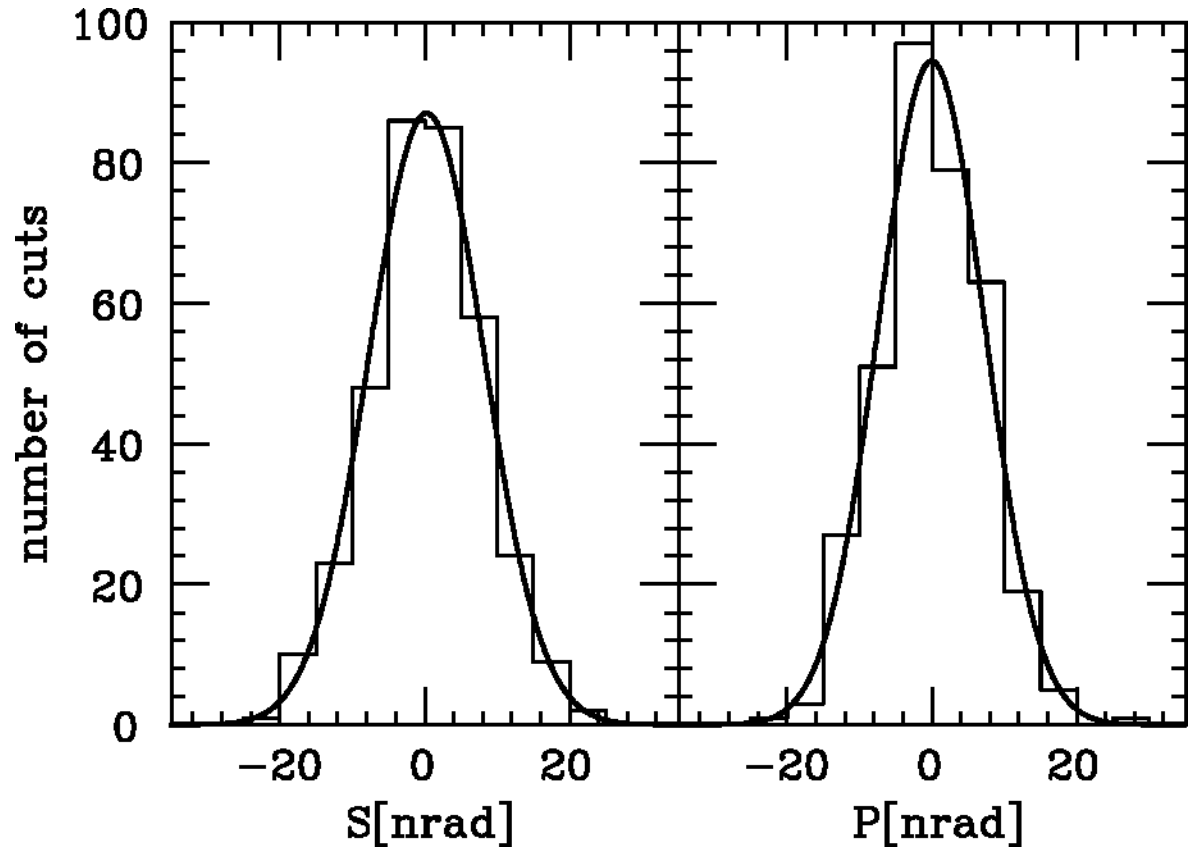

FIG. 15. 1997–1998 results for the signal, *S*, and quadrature, *P*, values for 346 pairs of data points with opposite *orientations* of the composition dipole on the pendulum tray. The smooth curves are Gaussian fits to the histograms.

25 $\mu$rad and 50 $\mu$rad, respectively.

(5) The temperature-control system for the air inside the thermo-acoustic shield was altered in an unsuccessful attempt to improve the stability of the air temperature.

The 1997–1998 data comprise a total of 692 cuts containing 19 changes in the *orientation* of the composition dipole. The $s_i$ and $p_i$ values from 20 data sets containing 346 pairs of cuts with opposite *orientations* of the composition dipole were were combined, using inverse error-squared weighting, to yield

$$S = +0.31 \pm 0.41 \text{ nrad}$$

$$P = -0.34 \pm 0.37 \text{ nrad}, \tag{16}$$

where the errors are statistical only. Confidence intervals were again established by binning all 346 $s_i$ and $q_i$ values and fitting the resulting histograms with Gaussians as shown in Fig. 15. The Gaussian fits to the *S* and *P* histograms were excellent, with $\chi^2$ probabilities of 99% and 12%, respectively. The the best-fit *S* and *P* values,

$$\langle S \rangle = +0.26 \pm 0.43 \text{ nrad}$$

$$\langle P \rangle = -0.04 \pm 0.39 \text{ nrad}, \tag{17}$$

agree well with those given in Eqs. (16).

### F. Statistical errors and thermal noise

Our statistical errors are only slightly greater than the values expected from fluctuations in the thermal energy in the pendulum. Figure 16 shows Fourier spectra of the $\theta$ signal taken when the attractor was rotating and when it was stationary. The smooth curve shows a first-principles calculation of the spectral density of thermal noise in $\theta$ for a pendulum whose quality factor $Q_0$ is dominated by internal friction in the suspension fiber [26],

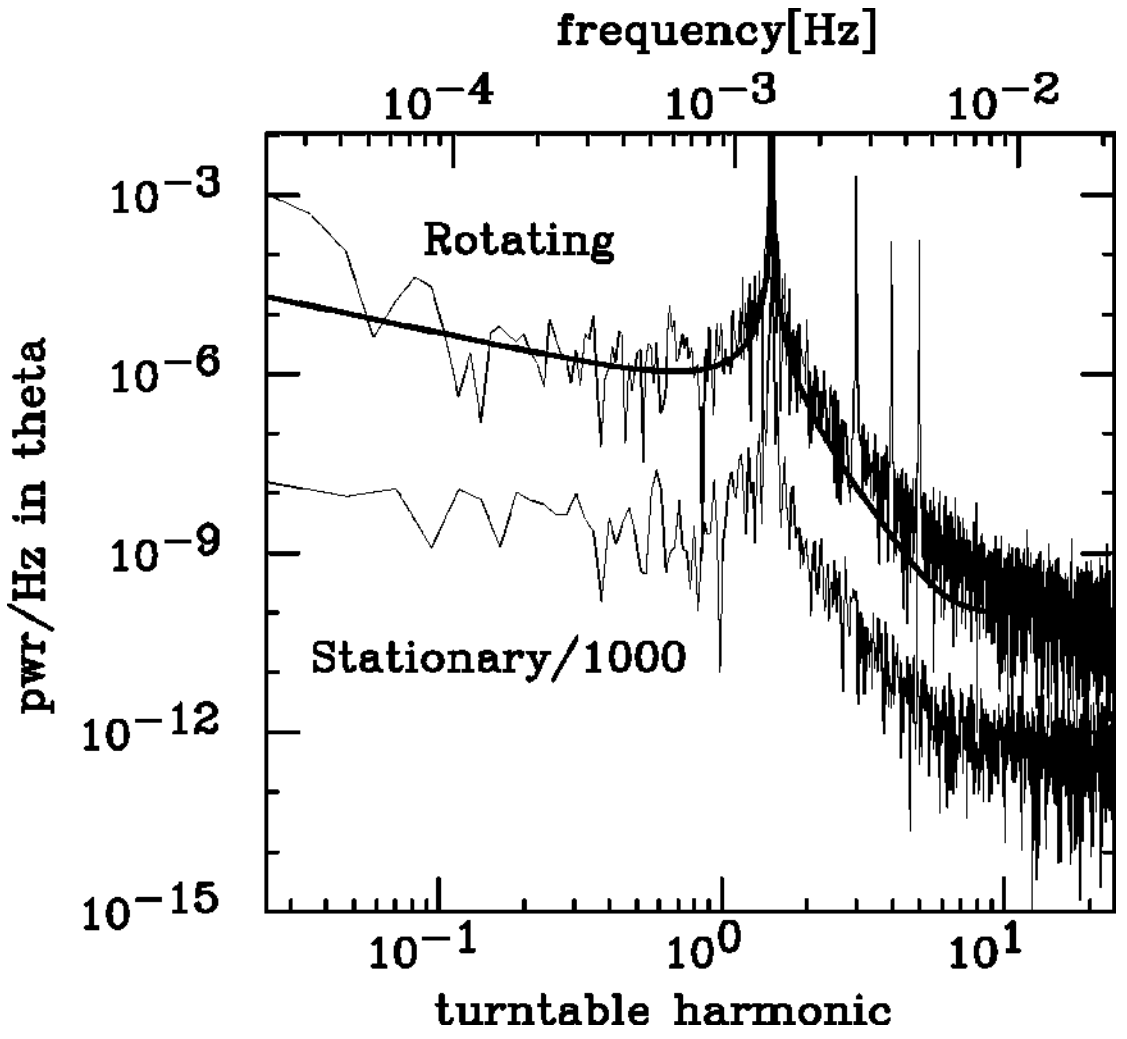

FIG. 16. Spectral power density of the autocollimator angle output. The lower spectrum (which is shifted down by 3 decades) was taken with a stationary attractor. The prominent peak at 1.3 mHz is the free resonance of the pendulum. The upper curve, taken with a rotating attractor, shows gravity-gradient signals at the 3rd, 4th and 5th harmonics of the revolution frequency. The smooth curve shows the thermal-noise prediction of Eq. (18).

$$\Theta^2(\omega) = \left[ \frac{4k_BT}{\kappa \omega Q_0} \left[ \left( 1 - \frac{\omega^2}{\omega_0^2} \right)^2 + \left( \frac{1}{Q_0} \right)^2 \right]^{-1} + \delta\theta^2 \right] \times \left[ \frac{1}{1+(\omega \tau_E)^2} \right]^2, \tag{18}$$

where $T$ is the temperature, $k_B$ is the Boltzmann constant, $\omega_0$ and $\omega$ are the free-resonance and signal frequencies of the pendulum, $\delta\theta^2$ is the density of the (white) autocollimator noise, and $\tau_E$ is the electronic time constant of the 2-pole filter in the autocollimator electronics. It is worth noting that the results in Fig. 16 are not consistent with velocity damping; this is expected because gas and eddy-current damping mechanisms are negligible in our instrument.

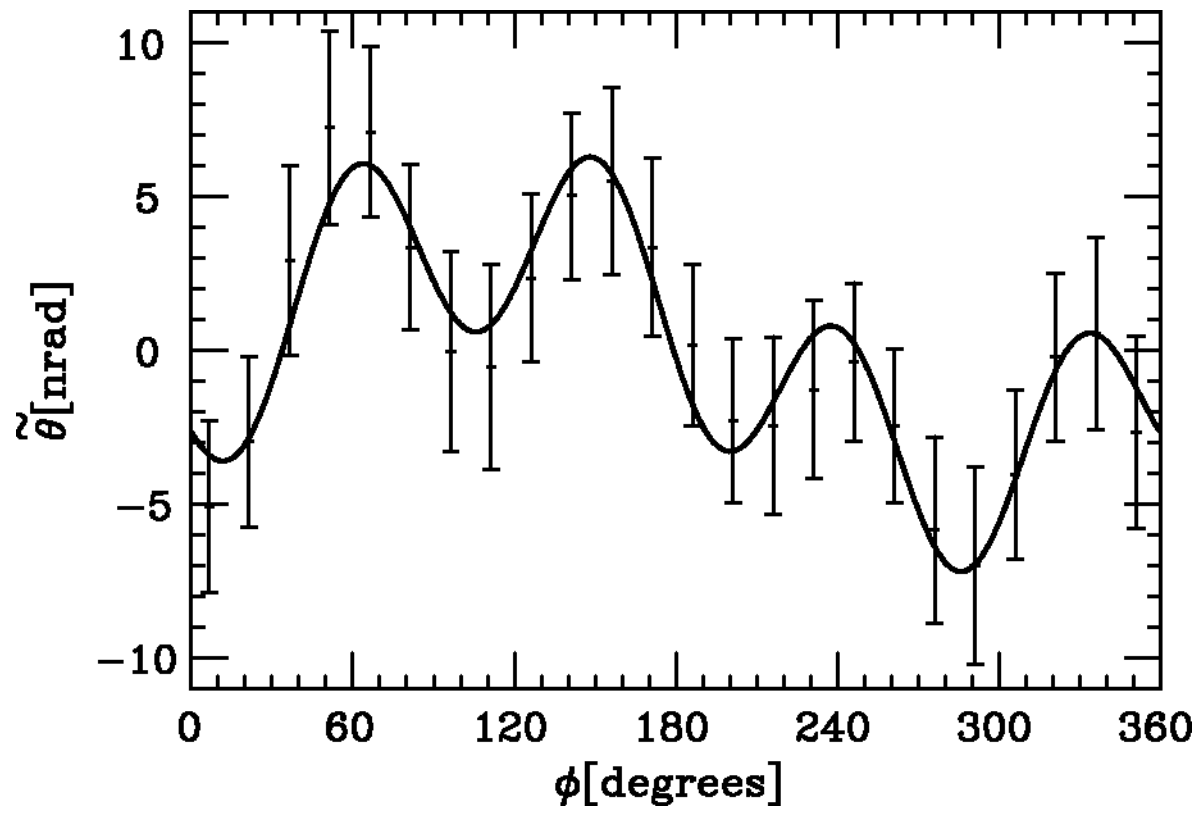

FIG. 17. Weak $1\omega$ and $4\omega$ signals produced by loading the pendulum with $q_{31}$ bodies and removing a 1.4 kg Pb block from the attractor. Results from 21 days of data were binned in $\phi$ modulo 360°. The error bars are derived from the scatter of the individual values in each bin. The smooth curve shows the best fit to the data. The extracted $m=1$ and $m=4$ amplitudes agreed with the expected gravitational $l=3$ and $l=4$ torques.

TABLE VII. $1\sigma$ systematic errors in the 1996 $S$ and $P$ values.

| Driving term | Normal magnitude | Error (nrad) |
|---|---|---|
| $l=2,3$ gravity-gradient correction[a] | 2.5 nrad | $\pm 0.03$ |
| $l=4$ gravity gradient | | $\pm 0.18$ |
| $1\omega$ air temperature fluctuations | $0.03\pm 0.01$ mK | $\pm 0.04$ |
| $1\omega$ shield temperature fluctuations | $0.03\pm 0.01$ mK | $\pm 0.05$ |
| $1\omega$ temperature gradients | $133\pm 2$ $\mu$K/cm | $\pm 0.04$ |
| $1\omega$ magnetic fields | 0.24 mG | $\pm 0.01$ |
| $1\omega$ floor tilt | $0.42\pm 0.42$ nrad | $\pm 0.03$ |
| Total | | $\pm 0.20$ |

[a]The $l=2$ and $l=3$ corrections had magnitudes of 2.5 and 2.4 nrad, respectively; because their phases differed, the total $l=2+l=3$ correction was less than the sum of the magnitudes.

Equation (18) shows that the signal-to-noise ratio improves with increasing signal frequency until, due to attenuation of the signal from pendulum inertia and electronic time constants, the autocollimator noise becomes significant. We operated with a signal frequency $\omega=2/3\omega_0$. With this choice, each 6-cycle cut had the same relative phases between the free torsional oscillations and the signal. The optimum signal-to-noise ratio of the $1\omega$ signal actually occurred for $\omega>\omega_0$ but the attenuation of the higher harmonics (which provided useful diagnositics) would be much larger at such frequencies.

### G. Demonstration of experimental sensitivity

We demonstrated the sensitivity and precision of our instrument by imposing feeble gravitational torques on the pendulum. These were produced by placing the $q_{31}$ gradiometer bodies on the pendulum and removing a 1.4 kg Pb block from the midplane of the attractor 49.3 cm from the suspension fiber. We calculated that the $l=3$ and and $l=4$ gravitational couplings of the Pb block to the pendulum's $q_{31}$ and $q_{44}$ moments would produce $1\omega$ and $4\omega$ signals of 3.2 and 4.6 nrad, respectively. The $1\omega$ torque, which had the same frequency as a UFF-violating signal, simulated a differential acceleration signal of $\Delta a=2.4\times 10^{-12}$ cm/s$^2$. The net results from data with and without the Pb block are shown in Fig. 17. The extracted signals, $a_1=3.6\pm 1.4$ and $a_4=3.5\pm 1.2$ nrad, agreed with our prediction. The measured value of the simulated UFF-violating effect was $\Delta a=(2.9\pm 1.1)\times 10^{-12}$ cm/s$^2$.

## IV. SYSTEMATIC EFFECTS

### A. Strategy

Systematic effects were studied by identifying possible driving terms (gravity gradients, magnetism, temperature variations and gradients, etc.) and making separate measurements in which, one at a time, each of these driving terms was deliberately introduced at a large enough level to cause a perceptible twist. The ratio of induced twist to driving term magnitude was our sensitivity to that driving term. The systematic effects were found by multiplying these sensitivities by the measured values of the driving terms under normal operating conditions. Our signal, defined in Eq. (12), was constructed so that a false signal could only occur if the driving terms or sensitivities were *configuration* dependent, i.e. differed for the two test-body *configurations* as well as for the two pendulum *angles*.

Gravitation was the only systematic effect large enough to warrant a correction to the UFF signal. The uncertainty in the gravitational correction and the upper limits on other systematic effects (see Tables VII and VIII) were summed in quadrature to obtain our total systematic error.

TABLE VIII. $1\sigma$ systematic errors in the 1997–1998 $S$ and $P$ values.

| Driving term | Normal magnitude | Error (nrad) |
|---|---|---|
| $l=2,3$ gravity-gradient correction[a] | 1.2 nrad | $\pm 0.01$ |
| $l=4$ gravity gradient correction | 0.15 nrad | $\pm 0.02$ |
| $1\omega$ air temperature fluctuations | $0.04\pm 0.02$ mK | $\pm 0.05$ |
| $1\omega$ shield temperature fluctuations | $0.08\pm 0.04$ mK | $\pm 0.07$ |
| $1\omega$ temperature gradients | $61.5\pm 2.5$ $\mu$K/cm | $\pm 0.02$ |
| $1\omega$ magnetic fields | 0.24 mG | $\pm 0.01$ |
| $1\omega$ floor tilt | $0.39\pm 0.44$ nrad | $\pm 0.03$ |
| Total | | $\pm 0.10$ |

[a]The $l=2$ and $l=3$ corrections had magnitudes of 0.1 and 1.2 nrad, respectively; because their phases differed, the total $l=2+l=3$ correction was less than the sum of the magnitudes.

### B. Gravitational effects

Two separate gravitational effects were involved: direct gravity-gradient torques described by Eq. (6) and ''sidepull'' effects arising from the gravitational deflection of the pendulum toward the attractor.

#### *1. Gravity-gradient torques*

The dominant gravity-gradient systematic error in $S$ and $P$ came from small variations in re-positioning the test bodies on the pendulum tray when test bodies were interchanged. The gravity-gradient corrections were proportional to the *configuration*-dependent pendulum moments in Table VI times the attractor fields given in Table IV. The root-mean-square gravity-gradient corrections applied to the individual data sets in the 1996 (1997–1998) experiments were 7.1 nrad (2.7 nrad), while the net corrections to the 1996 and 1997–1998 data sets were 2.5 nrad and 1.2 nrad, respectively. (The net corrections are not simply based on the product of the average quantities in Tables VI and IV because the stray moments varied between *configuration* changes.)

We made a careful test of the accuracy of our gravity-gradient correction procedure. We produced a pendulum with deliberately exaggerated $q_{21}$ and $q_{31}$ moments by loading the tray with two $q_{21}$ bodies on one diagonal and two $q_{31}$ bodies on the opposite diagonal. The attractor was adjusted to produce exaggerated $Q_{21}$ and $Q_{31}$ gradients by rotating its center and upper segments away from their normal positions by 90° and 0.6°, respectively. We observed a $1\omega$ pendulum twist of $\tilde{a}_1 = -8.14 \pm 0.06$ $\mu$rad, $\tilde{b}_1 = -20.13 \pm 0.06$ $\mu$rad. We then carried out our usual procedure for measuring the $q_{21}$ and $q_{31}$ moments of the pendulum and the $Q_{21}$ and $Q_{31}$ gradients of the attractor and computed the gravity-gradient correction. The calculated gravity-gradient torque, shown in Fig. 18, agreed to better than 0.7% with the observed twist. We therefore assigned a 1% systematic uncertainty to the $l=2$ and $l=3$ gravity-gradient corrections. The torque due to $l=4$ gradients, 0.05 $\mu$rad, was calculated from the designed-in $q_{41}$ moment of the $q_{21}$ gradiometer bodies and the designed-in $Q_{41}$ moment of the attractor. We included this entire torque as a systematic error in the test of the correction procedure.

#### *2. Sidepull*

The ''sidepull'' deflection angle, $\gamma = 94$ nrad, of the pendulum was calculated from the expression

$$\gamma = \frac{g'}{g} = -\sqrt{\frac{8\pi}{3}}\frac{G}{g} Q_{11}, \qquad (19)$$

where $Q_{11} = -481$ g/cm$^2$, $G$ is Newton's gravitational constant and $g = 980$ cm/s$^2$ is the gravitational acceleration on the Earth's surface. The sidepull deflection flexes the upper attachment of the prehanger fiber and so can lead to a spurious pendulum twist by the well-known ''tilt effect.'' The twist arising from the sidepull deflection was measured by stopping the attractor turntable and using the 3 air suspension legs to tilt the torsion balance. The legs were energized so that the tilt axis rotated in the horizontal plane at the same frequency as the actual sidepull under normal operation. We found (see Fig. 19) that for fiber 1 (fiber 2) a 103 $\mu$rad rotating tilt induced a 5.5 $\mu$rad (8.0 $\mu$rad) sinusoidal twist for a sensitivity of 0.0535 (0.0775). Thus, sidepull induced a pendulum twist of 5.03 nrad (7.29 nrad). But because the sidepull deflection was independent of the dipole *orientation*, sidepull had no effect on our signal.

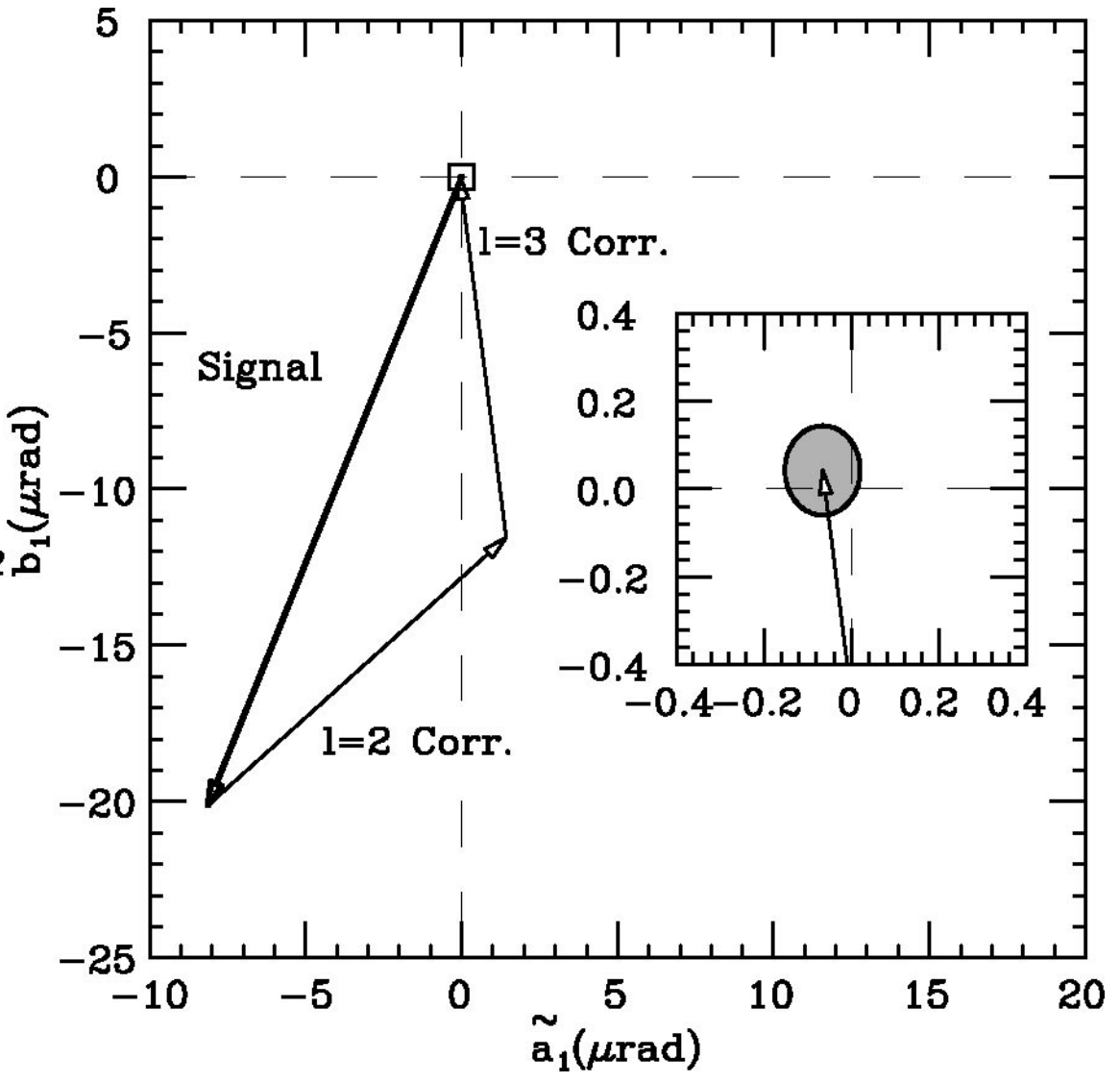

FIG. 18. Test of the gravity gradient correction. The heavy arrow shows the measured $1\omega$ signal when large $l=2$ and $l=3$ moments were deliberately introduced into both the pendulum and the attractor. The lighter arrows show the computed corrections for $l=2$ and $l=3$ gravity gradients. The inset, a magnified view of the origin, shows the accuracy of the corrections. The shaded region is the $1\sigma$ error ellipse of the corrected signal.

#### *3. Crank effect*

If the fiber mounts on the damper plate were misaligned so that its center of mass did not lie exactly on the line connecting the mounts, the prehanger fiber and the main fi-ber would not be coaxial, forming a ''crank'' that allows the sidepull force on the pendulum to exert a torque on the prehanger fiber. This torque can easily be large enough to give an appreciable twist to the prehanger fiber even though its torsion constant is quite high. If the sidepull acceleration had the same values at the sites of the damper and the pendulum, the crank effect would have been included in the table-tilt measurements discussed above. However, the attractor's horizontal field at the damper site is much smaller than it is at the pendulum, and the effect cannot be checked by tilting apparatus (which simulates a uniform sidepull field).

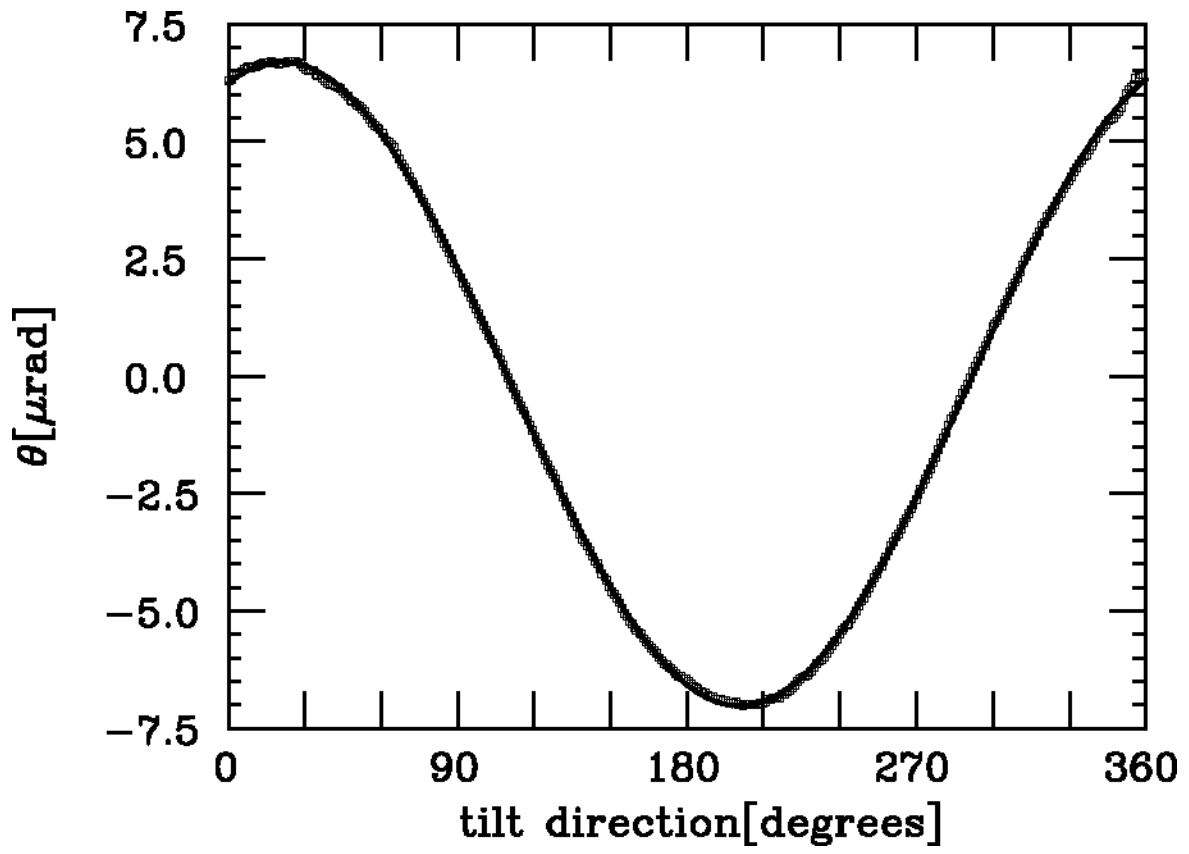

FIG. 19. Measurement of the tilt-twist coupling. The 8.0 $\pm 0.1$ $\mu$rad pendulum twist was caused by imposing a rotating tilt of 103 $\mu$rad. The induced twist was well described by a sine function.

If the horizontal acceleration due to the attractor at the site of the damper can be neglected, the misalignment torque on the prehanger fiber is

$$T_{\mathrm{p}}=m_{\mathrm{d}}\epsilon g'=\kappa_{\mathrm{p}}\theta, \tag{20}$$

where $m_{\mathrm{d}}=3.7$ g is the damper mass, $\epsilon$ is the horizontal misalignment between the fiber attachments and the damper center of mass, $g'=9.19\times10^{-5}$ cm/s$^2$, is the gravitational acceleration of the pendulum toward the attractor, and $\kappa_{\mathrm{p}}=250$ erg/rad is the torsion constant of the prehanger fiber. A misalignment of 0.01 mm would twist the prehanger, and therefore the pendulum, by 1.4 nrad. This twist is small only because we had a relatively large-diameter prehanger fiber and a low-mass damper plate. Because this effect, like sidepull itself, is *configuration* independent, it did not affect the signal. The good agreement of the *configuration*-independent but *angle*-dependent signals with the expected sidepull effect indicates that that the crank effect could not have been as large as 1 nrad.

### C. Magnetic effects

A magnetic field that varied at the $1\omega$ frequency of the turntable could lead to a systematic error if the pendulum or any part of its suspension system (such as the damper disk) had a ferromagnetic contaminant. To affect our signal, the field itself or the magnetic moment of the pendulum-suspension system would have to be *configuration* dependent. We measured the $1\omega$ magnetic fields after every *configuration* change by removing the vacuum vessel and magnetic shielding and placing a three-axis flux-gate magnetometer at the usual locations of the pendulum and damper disk. The $1\omega$ fields at the pendulum and damper disk were $0.24\pm0.01$ mG and $0.016\pm0.010$ mG, respectively. As expected, the fields were independent of *configuration* and remained constant during the course of the experiment.

We determined the pendulum's magnetic moment after every *configuration* change by applying a magnetic field that varied at the turntable frequency and measuring the induced $1\omega$ twist. For the fiber-1 data, the field was produced by 2 permanent magnets 0.5 m from the center of the rotating turntable. These generated rotating $B$ fields of 1150 mG and $103\pm2$ mG at the locations of the pendulum and damper plate, corresponding to enhancement factors of about 6000 in both locations. The induced *configuration*-independent and *configuration*-dependent $1\omega$ twists were $29\pm16$ nrad and $43\pm16$ nrad, respectively. Assuming that the magnetic field enhancement factors were unchanged by the magnetic shielding, the magnetic systematic error for the fiber-1 data was $7\pm3$ prad.

For the fiber-2 data, we produced $\pm232$ mG time-varying fields at the pendulum location by modulating, at the normal attractor rotation frequency, the currents in a pair of 1 m diameter coils, 1.3 m from the turntable axis. The enhancement factor at the pendulum location was 1100; the factor at the damper location was not measured but is expected to be considerably larger. The induced *configuration*-independent and *configuration*-dependent $1\omega$ twist amplitudes were $15\pm5$ nrad and $7\pm4$ nrad. If we make the conservative assumption that any magnetic impurity is in the pendulum rather than the damper plate, the magnetic systematic error in the fiber-2 data is $7\pm4$ prad.

### D. Thermal effects

The equilibrium twist of the suspension fiber, the electronics and optics of the autocollimator, and the rotation stage at the uppermost fiber support are all temperature-sensitive in some degree. Two different effects must be considered: a $1\omega$ modulation of the average temperature and a rotating temperature gradient, perhaps caused by a temperature difference between the uranium attractor and the pseudo-attractor. However, these effects would produce an error only if they differed in the two test body *configurations* as well as in the two *angles* of the pendulum body.

#### 1. Modulation of the average temperature

This effect was studied by keeping the attractor at a fixed position and modulating instrument temperatures at the 0.9 mHz signal frequency. We modulated the air temperature within the thermal-acoustic shield and, independently, the temperature of the innermost active heat shield. A $23.7\pm0.1$ mK modulation of the air temperature typically induced a $1\omega$ twist of $37.9\pm3.2$ nrad, with a phase of $-1°\pm5°$ with respect to the temperature signal. As expected, this twist was nearly the same, $38.3\pm3.5$ nrad with a phase of $-15°\pm6°$, when the pendulum was rotated by 180°. The resulting sensitivity was $1.6\pm0.1$ nrad/mK. In the 1996 (1997–1998) UFF data, the same sensor measured air-temperature variations that followed the dipole of $0.03\pm0.01$ mK with a phase of $304°\pm36°$ $(0.04\pm0.02$ mK with a phase of $166°\pm15°)$, where these phases are with respect to the attractor angle. This gave an air-temperature-modulation systematic error in $S$ and $P$ of $0.04\pm0.02$ nrad $(0.05\pm0.01$ nrad) respectively.

A $70\pm1$ mK modulation of the innermost heat shield induced a $132.0\pm3.4$ nrad $1\omega$ twist. During the 1996 (1997–1998 data) the same temperature sensor recorded $1\omega$ variations that followed the dipole of $0.03\pm0.01$ mK with a phase of $135°\pm20°$ $(0.08\pm0.04$ mK with a phase of $180°\pm21°)$, leading to shield-temperature systematic errors in the 1996 (1997–1998) $S$ and $P$ values of $0.05\pm0.01$ nrad $(0.07\pm0.01$ nrad$)$.

#### 2. Rotating temperature gradients

A rotating temperature gradient could have arisen from radioactive decay of the attractor or, if thermal equilibrium had not been reached, from the different heat capacities of the attractor and the pseudo-attractor. The $^{238}$U attractor was a source of $\alpha$, $\beta$, and $\gamma$ radiation, almost all of which was absorbed within the attractor itself. The disintegration of a $^{238}$U nucleus via the decay chain $^{238}\mathrm{U}\rightarrow{}^{234}\mathrm{Th}\rightarrow{}^{234}\mathrm{Pa}\rightarrow{}^{234}\mathrm{U}$ releases 6.74 MeV. The remainder of the decay chain may be ignored as the $^{234}$U half-life is $2.5\times10^5$ yrs, and the highly enriched material initially contained very little $^{234}$U. Therefore, radioactivity generated only 25 mW of heat, where we have included the fact that neutrinos do not heat the attractor.

Our sensitivity to rotating temperature gradients was tested by removing a part of the pseudo-attractor and replacing it with a resistor dissipating 60 W. When this setup was rotated at the usual rate, a pair of temperature sensors mounted on (but thermally insulated from) the underside of the granite table, $\sim$10 cm from the turntable axis, recorded a $405\pm4$ mK $1\omega$ modulation. The corresponding $1\omega$ twist was $4.5\pm5.6$ nrad, yielding a sensitivity of $0.011\pm0.014$ nrad/mK. During normal data taking, the largest *configuration*-dependent $1\omega$ signal ever seen on the same pair of sensors was 2.82 mK, which corresponds to a $0.03\pm0.04$ nrad twist of the pendulum. This is a conservative upper limit on the associated systematic error because the *configuration*-dependent effect should be a small fraction of the *configuration*-independent twist measured here.

### E. Attractor imbalance

An imbalance of the attractor turntable could have tilted the torsion balance at the signal frequency and produced a systematic error that was not cancelled by the *configuration* changes. The $1\omega$ signal from a two-axis electronic level mounted on the torsion balance, which never exceeded 12 nrad, was multiplied by the tilt sensitivity given in Sec. IV B 2 to yield the upper limits on attractor imbalance systematic errors given in Tables VII and VIII.

### F. Internal consistency checks

The UFF-violating signal defined in Eq. (12), the *configuration* and-*angle*-dependent combination of the $a_1$ and $b_1$ coefficients from the 4 experimental states defined by *configurations* $\mathcal{A}$ and $\mathcal{B}$ and *angles* $\mathcal{N}$ and $\mathcal{R}$, cancels a large number of potential systematic errors. The other 3 independent signals that can be formed from the 4 states may be of some interest as well. These diagnostic signals are completely insensitive to UFF violation, but are very sensitive to systematic effects that subtract in the UFF-violating signal.

The *configuration*-independent but *angle*-dependent version of the signal given in Eq. (12) does not cancel the sidepull effect. This diagnostic signal had magnitudes (i.e. the quadrature sums of the corresponding ‘‘$S$’’ and ‘‘$P$’’ coefficients) and statistical errors in the 1996 (1997–1998) data of $4.12\pm1.10$ nrad ($7.48\pm0.41$ nrad) that were entirely consistent with the expected sidepull effect. The *configuration*- and-*angle*-independent combination of the data, which does not cancel false effects such as ‘‘rumble’’ in the turntable bearing, or overall temperature changes that varied at the turntable frequency, had magnitudes and statistical errors of $1.7\pm1.1$ nrad ($2.1\pm0.4$ nrad) in the 1996 (1997–1998) data. The *configuration*-dependent but *angle*-independent combination had magnitudes of $1.3\pm1.1$ nrad ($0.5\pm0.4$ nrad) in the 1996 (1997–1998) data. This diagnostic signal was much more sensitive than the UFF-violating signal to long-term drifts in systematic effects and calibrations because *angles* were changed much more frequently than *configurations*.

The agreement of the first diagnostic signal with expections based on the measured sidepull effect, and the small size of the second and third diagnostic signals, gave us confidence that potential systematic effects in the UFF-violating signal had canceled to a satisfactory level.

## V. UFF-VIOLATING DIFFERENTIAL ACCELERATION

We combine our 1996 and 1997–1998 values of $S$ and $P$ given in Eqs. (14) and (16), and the systematic errors from Tables VII and VIII, to obtain

$$S=+0.14\pm0.36\pm0.12 \text{ nrad}$$

$$P=-0.06\pm0.33\pm0.12 \text{ nrad}, \tag{21}$$

where the first errors are statistical and the second systematic. The central values and systematic errors for the combined data were obtained by weighting the 1996 and 1997–1998 values by their inverse squared statistical errors, while the statistical error was obtained in the usual way. Our procedure for evaluating the total systematic error is conservative; it assumed that the systematic errors in the two data sets were perfectly correlated, whereas the systematic errors in the two sets were often limited by statistical uncertainties in measurements of the ‘‘driving terms.’’

As both $S$ and $P$ of the combined data are consistent with zero, we use the results in Eq. (21) to obtain our final value for the differential acceleration of Cu and Pb toward $^{238}$U:

$$\Delta a\equiv a_{\mathrm{Cu}}-a_{\mathrm{Pb}}=\kappa S/(Md)=(1.0\pm2.6\pm0.9)\times10^{-13} \text{ cm/s}^2, \tag{22}$$

where both errors are $1\sigma$. The corresponding 68% and 95% confidence limits on $|\Delta a|$ are $2.9\times10^{-13}$ cm/s$^2$ and $5.7\times10^{-13}$ cm/s$^2$ respectively, where we have combined the statistical and systematic errors in quadrature and found the symmetric region containing 68% and 95% of the Gaussian probablilty density.

## VI. CONSTRAINTS ON NEW PHYSICAL INTERACTIONS

### A. Equivalence-principle violating Yukawa interactions

We now use the result in Eq. (22) to constrain possible equivalence-principle violating interactions as functions of the $\lambda$. To do this, we must include the effect of the attractor's modulation of local vertical that was ignored in Eqs. (2) and (5). The torque on the torsion fiber can be written as

$$T=\frac{\vec{F}_1\times\vec{F}_2\cdot\vec{d}}{|\vec{F}_1+\vec{F}_2|}, \tag{23}$$

TABLE IX. $I_{\rm E}(\lambda)$ for our 13-layer model of the Earth.

| $\lambda$ (m) | $I_{\rm E}(\lambda)$ |
|---|---|
| 10 | $3.94\times10^{-7}$ |
| $10^2$ | $4.31\times10^{-6}$ |
| $10^3$ | $4.69\times10^{-5}$ |
| $10^4$ | $9.81\times10^{-4}$ |
| $10^5$ | $1.36\times10^{-2}$ |
| $10^6$ | $1.72\times10^{-1}$ |
| $10^7$ | $8.77\times10^{-1}$ |
| $10^8$ | $9.98\times10^{-1}$ |

where $\vec{F}_1$ and $\vec{F}_2$ are the net forces on test bodies of compositions 1 and 2 respectively, and $\vec{d}$ is a vector that points from the center of test bodies 1 to the center of test bodies 2. These forces consist of the inertial force plus the gravitational and Yukawa interactions of the test bodies with the attractor and the entire Earth,

$$\vec{F}_i=M_i\left\{\vec{a}_c+\vec{g}\left[1+I_E\left(\frac{q}{\mu}\right)_i\left(\frac{q}{\mu}\right)_{\rm E}\right]+\vec{g}'\left[1+I_{\rm A}\left(\frac{q}{\mu}\right)_i\left(\frac{q}{\mu}\right)_{\rm A}\right]\right\}, \tag{24}$$

where the subscripts ''E'' and ''A'' refer to the Earth and attractor, $\vec{g}$ and $\vec{g}'$ are the gravitational accelerations toward the Earth and attractor, and $\vec{a}_c$ is the centrifugal acceleration of the Earth's rotation. The integral $I_{\rm A}(\lambda)$ is defined in Eq. (5). If the Earth were a uniform-density sphere, its integral would be [27]

$$I_{\rm E}(\lambda)=3\left(\cosh x-\frac{\sinh x}{x}\right)\left(\frac{1+y}{x^2}+\frac{1}{x}\right)e^{-(x+y)}, \tag{25}$$

with $x=R_{\rm E}/\lambda$ and $y=h/\lambda$ where $h$ is the height of the apparatus above the Earth's surface. However, this approximation substantially overestimates $I_{\rm E}$ for $\lambda\ll R_{\rm E}$ because the density of the Earth's crust is well below the average density of the entire Earth. We evaluated $I_{\rm E}(\lambda)$ by approximating the Earth as a series of 13 spherical shells whose radii, thicknesses and densities were found with the aid of Ref. [28]. The resulting values of $I_{\rm E}$, for an apparatus 1 m above the surface, are shown in Table IX. Retaining only those terms in Eqs. (23) and (24) that vary as the attractor rotates, we find

$$\frac{\Delta a}{g'}=\frac{T_{\rm y}^{(1)}}{Mdg'}=\alpha\Delta\left(\frac{\tilde{q}}{\mu}\right)\left[I_{\rm A}(\lambda)\left(\frac{\tilde{q}}{\mu}\right)_{\rm A}-I_{\rm E}(\lambda)\left(\frac{\tilde{q}}{\mu}\right)_{\rm E}+\frac{a_c\cos\Theta}{g}I_{\rm A}(\lambda)\left(\frac{\tilde{q}}{\mu}\right)_{\rm A}\right]\sin\phi, \tag{26}$$

where $\Theta=47.66°$ is the laboratory latitude and $M$, $d$, and $\phi$ have the same meanings as in Eq. (4). Note that for infinite range, where $I(\lambda)=1$, the first two terms in Eq. (26) cancel if the attractor and Earth have the same $(\tilde{q}/\mu)$. The intrinsic sensitivity $\mathcal{S}=\Delta(\tilde{q}/\mu)\langle(\tilde{q}/\mu)_{\rm A}I_{\rm A}(\lambda)\rangle$ is given in Table X.

TABLE X. Intrinsic sensitivity $\mathcal{S}(\lambda)=\Delta(\tilde{q}/\mu)\langle(\tilde{q}/\mu)_{\rm A}I_{\rm A}(\lambda)\rangle$ of the Rot-Wash instrument for 3 possible choices for the Yukawa charge: $\tilde{q}=N+Z$, $\tilde{q}=N-Z$ and $\tilde{q}=B-L$. The angular-brackets denote a sum over the U attractor, the Pb counterweight and the Al ''pseudo-attractor.''

| $\lambda$ (m) | $\mathcal{S}(\tilde{q}=N+Z)$ | $\mathcal{S}(\tilde{q}=N-Z)$ | $\mathcal{S}(\tilde{q}=B-L)$ |
|---|---|---|---|
| 0.010 | $1.23\times10^{-8}$ | $-3.78\times10^{-7}$ | $-4.65\times10^{-7}$ |
| 0.014 | $1.75\times10^{-7}$ | $-5.31\times10^{-6}$ | $-6.62\times10^{-6}$ |
| 0.020 | $1.53\times10^{-6}$ | $-4.57\times10^{-5}$ | $-5.77\times10^{-5}$ |
| 0.028 | $7.46\times10^{-6}$ | $-2.21\times10^{-4}$ | $-2.81\times10^{-4}$ |
| 0.05 | $5.45\times10^{-5}$ | $-1.60\times10^{-3}$ | $-2.05\times10^{-3}$ |
| 0.07 | $1.25\times10^{-4}$ | $-3.67\times10^{-3}$ | $-4.71\times10^{-3}$ |
| 0.10 | $2.47\times10^{-4}$ | $-7.23\times10^{-3}$ | $-9.29\times10^{-3}$ |
| 0.20 | $5.61\times10^{-4}$ | $-1.63\times10^{-2}$ | $-2.11\times10^{-2}$ |
| 0.50 | $8.44\times10^{-4}$ | $-2.46\times10^{-2}$ | $-3.17\times10^{-2}$ |
| 1.0 | $9.19\times10^{-4}$ | $-2.68\times10^{-2}$ | $-3.45\times10^{-2}$ |
| 2.0 | $9.41\times10^{-4}$ | $-2.74\times10^{-2}$ | $-3.53\times10^{-2}$ |
| 5.0 | $9.48\times10^{-4}$ | $-2.76\times10^{-2}$ | $-3.56\times10^{-2}$ |
| 10.0 | $9.48\times10^{-4}$ | $-2.77\times10^{-2}$ | $-3.56\times10^{-2}$ |
| 20.0 | $9.49\times10^{-4}$ | $-2.77\times10^{-2}$ | $-3.56\times10^{-2}$ |
| 50.0 | $9.49\times10^{-4}$ | $-2.77\times10^{-2}$ | $-3.56\times10^{-2}$ |

Figure 20 shows our limits on the strength of hypothetical Yukawa interactions coupling to 3 possible charges: $\tilde{q}=B$, $\tilde{q}=B-L$, and $\tilde{q}=N-Z$, where $B=N+Z$ and $L=Z$ are the baryon and lepton numbers, respectively. All three of these charges have been suggested as possible sources of vector fields; $\tilde{q}=B-L$ is particularly interesting because this quantity is conserved in grand unified theories that do not conserve $B$ and $L$ individually. The constraints on $\alpha(\lambda)$ were computed from Eq. (26) using the result in Eq. (22). For $\tilde{q}=B$, the cancellation of the first two terms in Eq. (26) is almost complete, which weakens our $\tilde{q}=B$ constraints for $r\geqslant R_{\rm E}$.

Figure 20 shows that we improved the $\tilde{q}=B=N+Z$ constraints by a factor of 300 at $\lambda\approx0.5$ m, and by up to a factor of 100 for 10 km$\leqslant\lambda\leqslant$1000 km; the the tightest previous constraints in the latter region came from Galileo-type measurements [10,11]. We improved existing constraints [4,29–31] on interactions coupled to $\tilde{q}=N-Z$ by a factor of at least 680 for 2.3 cm$\leqslant\lambda\leqslant10^6$ m and ruled out an earlier suggestion [8] of such an interaction. Our constraints on interactions with $\tilde{q}=B-L$ improve on previous work [4,29,31] by a factor of at least 500 for 5.2 cm$\leqslant\lambda<1$ m and establish the tightest bounds up to $\lambda=420$ m.

Constraints for Yukawa interactions coupled to other possible charges can be obtained from Eq. (26) along with the attractor and Earth integrals given in Tables X and IX. Figure 21, which applies to Yukawa interactions mediated by vector particles, shows how our constraint varies with the parameter $\theta_5$ that specifies the most general vector charge of stable, electrically neutral matter [27], $\tilde{q}=B\cos\theta_5+L\sin\theta_5$. The $B/\mu$ and $L/\mu$ values for our test bodies and attractor are given in Table XI.

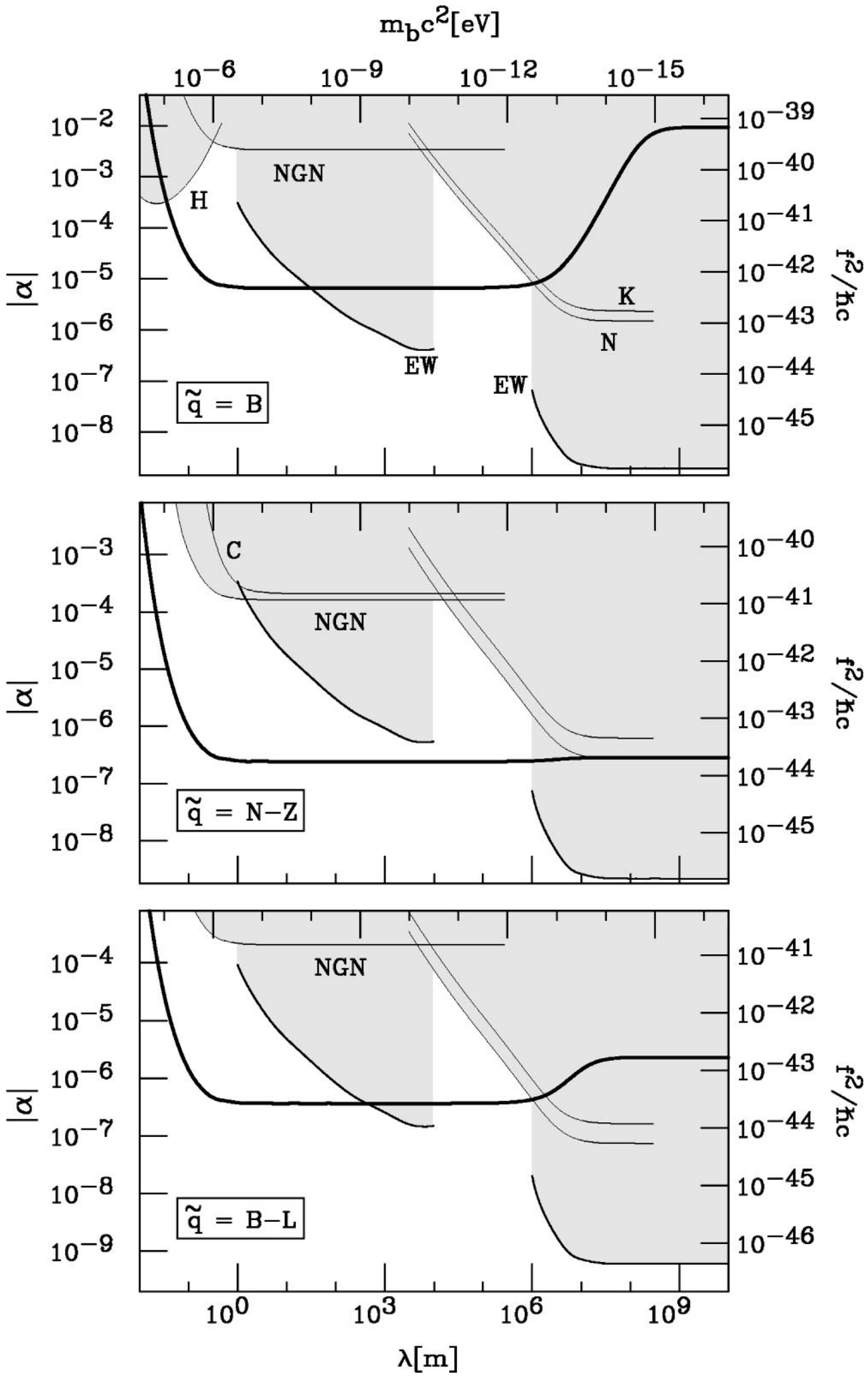

FIG. 20. 95% confidence limits on $|\alpha|$ vs λ for hypothetical interactions coupling to vector charges $\tilde{q}=B$, $\tilde{q}=N-Z$, or $\tilde{q}=B-L$. The heavy curves are from this work; the shaded region was excluded by previous results labeled as follows: EW [4], NGN [29], N [10], K [11], C [30], H [31]. Constraints from Galileo-type measurements were computed using the Earth model given in Table IX.

### B. Power-law interactions from 2-boson exchange

The simultaneous exchange of a pair of massless bosons between a pair of test bodies can lead to a static power-law potential even if the single-boson exchange potential vanishes because of a conservation law. The van der Waals interaction between neutral atoms is a familiar example. If the couplings of the massless bosons are composition dependent (as they are in essentially all scenarios), then the two-boson exchange interaction violates the UFF.

TABLE XI. Vector charge-to-mass ratios, $B/\mu$ and $L/\mu$, of the test bodies and attractor.

| | Cu | "Pb"[a] | $^{238}$U |
|---|---|---|---|
| $B/\mu$ | 1.0011166 | 1.0001694 | 0.9962151 |
| $L/\mu$ | 0.4563649 | 0.3975718 | 0.3850916 |

[a]Value for the actual test-body alloy.

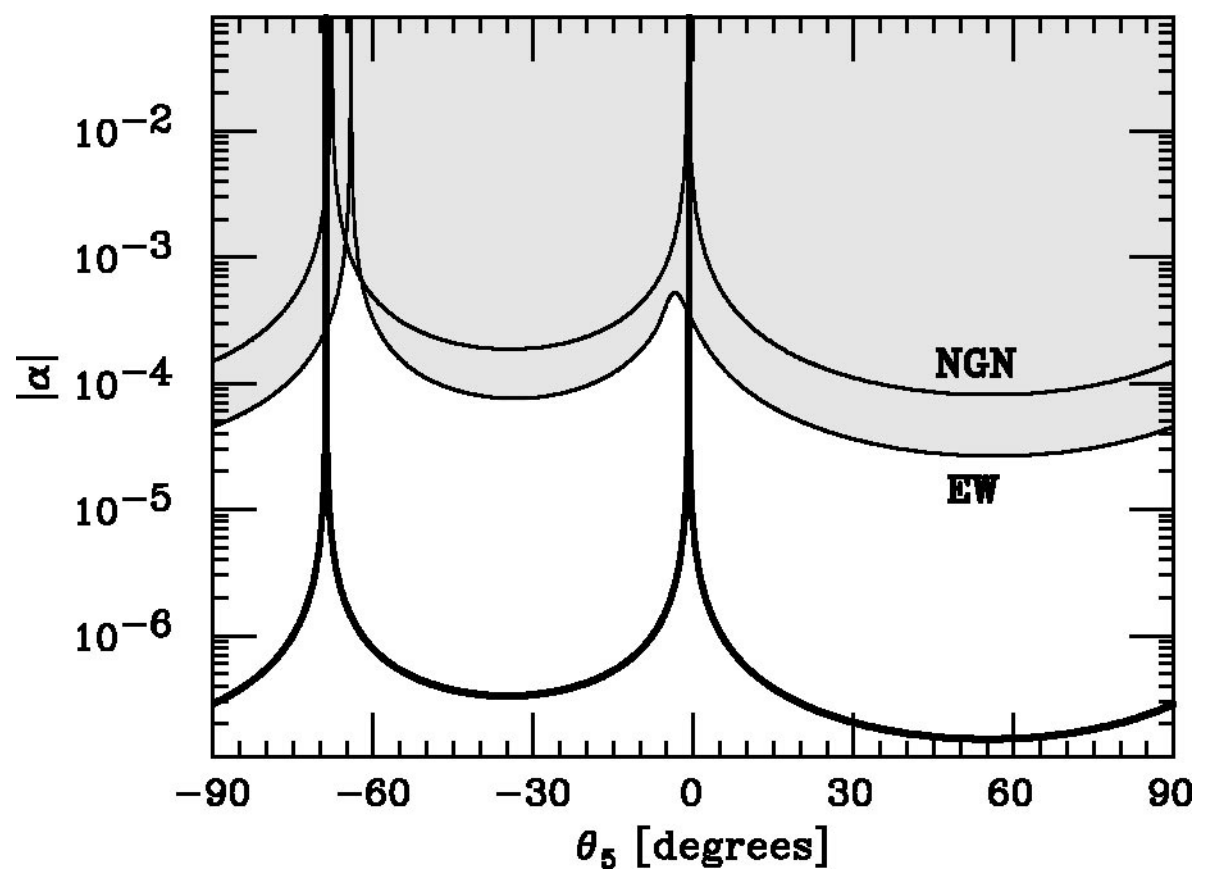

FIG. 21. 95% confidence limits on $\alpha$ as a function of the vector charge parameter, $\theta_5$, for a hypothetical Yukawa interaction with $\lambda=1$ m. The heavy curve is from this work; the shaded region was excluded by previous results. The curves labeled EW and NGN are from Refs. [4] and [29] respectively. The constraints from this work are valid for $1.0\ \mathrm{m}\leqslant\lambda\leqslant10^6$ m.

Feinberg and Sucher [32] have discussed power-law potentials that could arise from two-gluon exchange and considered experimental limits on potentials of the form

$$V_{12}(r)=\Lambda_N\frac{\hbar c}{r}\left(\frac{r_0}{r}\right)^{N-1}B_1B_2 \tag{27}$$

where $r_0=1$ fm and $B_1$ and $B_2$ are the baryon numbers of the interacting bodies. The potential in Eq. (27) leads to a differential acceleration

$$\Delta\vec{a}=\Lambda_N r_0^{N-1}\Delta\left(\frac{B}{\mu}\right)\langle B/\mu\rangle_{\mathrm{A}}\vec{J}_N, \tag{28}$$

where $M$ is a test body mass and

$$\vec{J}_N=\frac{N\hbar c}{4Mu^2}\int\rho_{\mathrm{T}}(\vec{r})\rho_{\mathrm{A}}(\vec{r}\,')\frac{\vec{r}-\vec{r}\,'}{|\vec{r}-\vec{r}\,'|^{N+2}}d^3\vec{r}d^3\vec{r}\,'; \tag{29}$$

$\rho_{\mathrm{T}}$ and $\rho_{\mathrm{A}}$ refer to the density distributions of all 4 test bodies and the attractor, respectively. We computed the $J_N$ integrals, listed in Table XII, by direct numerical integration over the attractor and pendulum geometries. Our results in Fig. 20 and Eq. (22), together with Table XII, yield substantially improved constraints on $N=2$ and $N=3$ power-law potentials. Table XIII shows our limits on such potentials, along with previous results given in Ref. [32].

Fischbach and Krause [33] have recently shown that limits on power-law potentials from UFF experiments also pro-vide powerful laboratory constraints on nucleon-spin-dependent couplings arising from the exchange of very-low-mass pseudoscalars. To lowest order, exchange of a massless pseudoscalar with $\gamma_5$ couplings leads to a spin-dependent potential

$$V_{12}^{(1)}(r,\vec{\sigma}_1,\vec{\sigma}_2)=\frac{g^2}{16\pi}\frac{\hbar^2}{M_N^2c^2}\frac{S_{12}}{r^3}, \tag{30}$$

where $g$ is a pseudoscalar coupling constant, $M_N$ is the nucleon mass, the nucleon spin is $\vec{s}=\vec{\sigma}\hbar/2$, and $S_{12}\equiv 3(\vec{\sigma}_1\cdot\hat{r})(\vec{\sigma}_2\cdot\hat{r})-(\vec{\sigma}_1\cdot\vec{\sigma}_2)$. The second-order process involving the exchange of a pair of pseudoscalars also leads to a $1/r^3$ spin-independent potential [33]

$$V_{12}^{(2)}(r)=-\frac{g^4}{64\pi^3}\frac{\hbar}{M_N^2c^3}\frac{1}{r^3}. \tag{31}$$

TABLE XII. $J_N$ integrals of the Rot-Wash instrument.

| $N$ | $J_N$ |
|---|---|
| 2 | $1.27\times10^{33}$ s$^{-2}$ |
| 3 | $9.1\times10^{31}$ cm$^{-1}$ s$^{-2}$ |
| 4 | $6.9\times10^{30}$ cm$^{-2}$ s$^{-2}$ |

TABLE XIII. 68% confidence limits on the strength of power-law potentials.

| $N$ | $\|\Lambda_N\|$ | |
|---|---|---|
| | Eöt-Wash group | Ref. [32] |
| 1 | $5.8\times10^{-48}$ [a] | $1\times10^{-47}$ |
| 2 | $2.4\times10^{-30}$ [b] | $1\times10^{-26}$ |
| 3 | $3.4\times10^{-16}$ [b] | $1\times10^{-12}$ |
| 4 | $4.5\times10^{-2}$ [b] | $3\times10^{-3}$ |

[a]Reference [4].
[b]This work.

The resulting differential acceleration is

$$\Delta\vec{a}=-\frac{1}{64\pi^3M_N^2c^4}\left[g_p^2\Delta\left(\frac{Z}{\mu}\right)+g_n^2\Delta\left(\frac{N}{\mu}\right)\right][g_p^2\langle Z/\mu\rangle_A + g_n^2\langle N/\mu\rangle_A]\vec{J}_3, \tag{32}$$

where $g_p$ and $g_n$ are the pseudoscalar couplings of protons and neutrons. Our constraints on $g_p^2$ and $g_n^2$, along with those derived [34] from tests [31] of the gravitational inverse-square law, are shown in Fig. 22. Astrophysical considerations regarding the cooling rates of stars [35] provide constraints that are much tighter, but more model dependent, than the laboratory limits given in Fig. 22.

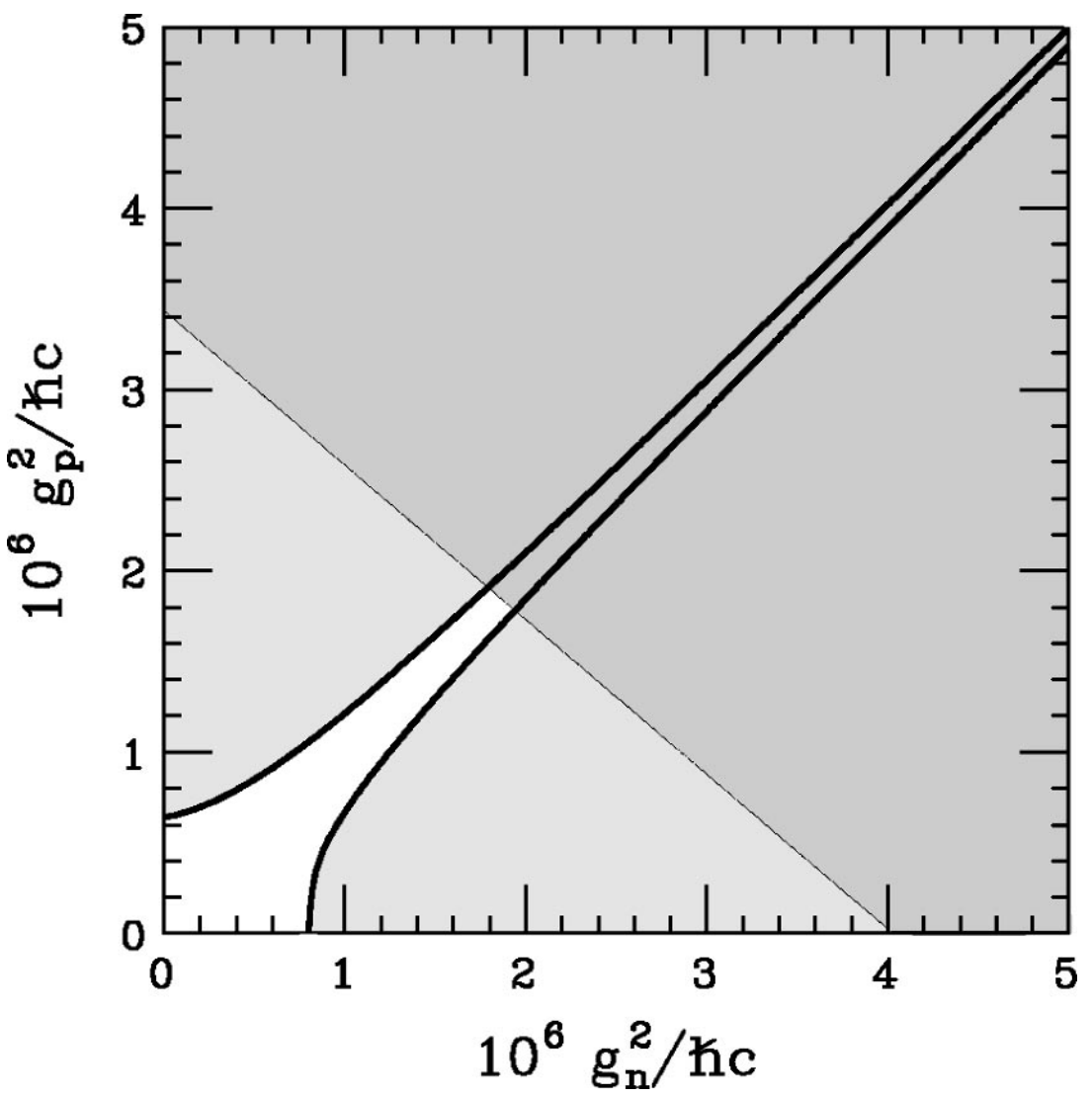

FIG. 22. 68% confidence limits on a hypothetical interaction between nucleons due to second-order exchange of pseudoscalars with $\gamma_5$ couplings; the excluded regions are shaded. The heavy solid curves are from this experiment; the light line, derived from the inverse-square tests of Ref. [31], is taken from Ref. [34]. Results apply for $\lambda\geqslant2$ m.

## VII. FUTURE PROSPECTS

This work has reached a differential acceleration sensitivity of $\sim3\times10^{-13}$ cm/s$^2$. If an object, initially at rest, had been given this acceleration approximately 2500 year ago and that acceleration had been maintained to this day, the object would now be moving as fast as the end of the minute hand on a standard wall clock. Improvements in this sensitivity, say by a factor of 10 or more, would require a large increase in the quality factor, $Q_0$, of the torsion oscillator as well as better control of systematic errors. Consequently, we have no plans at present for substantial improvements in the sensitivity of the Rot-Wash instrument. Instead, motivated the intense, continuing theoretical interest in the possibility that the ''extra'' spatial dimensions of supersymmetry could lead to violations of the gravitational Gauss law at the sub-millimeter scale [36], we are building an entirely new pendulum and attractor system that should have good sensitivity to violation of the $1/r^2$ law at length scales down to 0.1 mm.

## ACKNOWLEDGMENTS

This work was supported primarily by National Science Foundation Grants PHY-9104541 and PHY-9602494. The Department of Energy helped us acquire the high-purity depleted uranium, and provided some financial support. Professor W. F. Rogers designed the pendulum tray, Dr. David Sesko helped construct the instrument, and Dr. Steven Penn contributed to the autocollimator design. We are grateful for the assistance of several talented undergraduates. Tim Bast and Kurt Adelberger wrote programs for computing multipole moments of complex objects, Hans Vija and Jörn Häuser helped assemble the uranium attractor, and Kimberly Mauldin helped with torsion balance design.